\documentclass[10pt,superscriptaddress,twocolumn,aps,showpacs,nofootinbib,eqsecnum,twoside,prc]{revtex4}
\usepackage[pdfborder={0 0 0},colorlinks=true,linkcolor=blue,citecolor=blue,filecolor=blue,urlcolor=blue]{hyperref}   
\usepackage{url}
\usepackage{ulem}
\usepackage{slashed}
\usepackage{amssymb}
\usepackage{amsfonts}
\usepackage{mathbbol} 
\usepackage{amstext}
\usepackage{graphicx}
\usepackage{color}      
\usepackage{epsfig,amsmath,lscape}
\usepackage{soul}
\usepackage{mathptmx}                
\usepackage{dcolumn}                 
\usepackage{bm}                      
\usepackage[caption=false]{subfig}
\usepackage{subfig}

\begin{document}

\title{Examination of strangeness instabilities and effects of strange
  meson couplings in dense strange hadronic matter and compact stars}

\author{James R. Torres}
\affiliation{Departamento de F\'{\i}sica - CFM - Universidade Federal de Santa Catarina, \\ 
Florian\'opolis - SC - CP. 476 - CEP 88.040 - 900 - Brazil \\ email: james.r.torres@posgrad.ufsc.br}

\author{Francesca Gulminelli}
\affiliation{CNRS and ENSICAEN, UMR6534, LPC, \\ 
14050 Caen c\'edex, France \\ email:
gulminelli@lpccaen.in2p3.fr }

\author{D\'ebora P. Menezes}
\affiliation{Departamento de F\'{\i}sica - CFM - Universidade Federal de Santa Catarina, \\ 
Florian\'opolis - SC - CP. 476 - CEP 88.040 - 900 - Brazil \\ email: debora.p.m@ufsc.br}

\begin{abstract}
\begin{description}
\item[Background] : The emergence of hyperon degrees of freedom in neutron star matter has been associated to first order phase transitions in some phenomenological models, 
but conclusions on the possible physical existence of an instability in the strangeness sector are  strongly model dependent.
\item[Purpose] :  The purpose of the present study is to assess whether strangeness instabilities are related to specific values of the largely unconstrained hyperon interactions, and to study the effect of the strange meson couplings on phenomenological properties of neutron stars and supernova matter, once these latter are fixed to fulfill the constraints imposed by hypernuclear data. 
\item[Method] :  We  consider a phenomenological RMF model sufficiently simple to allow a complete exploration of the parameter space.  
\item[Results] We show that no instability at supersaturation density exists for the RMF model, as long as the parameter space is constrained by basic physical requirements. This is at variance with a non-relativistic functional, with a  functional behavior fitted through ab-initio calculations.
Once the study is extended to include the full octet, we show that the parameter space allows reasonable radii for canonical neutron stars as well as massive stars above two-solar mass, together with an important strangeness content of the order of 30\%, slightly decreasing with increasing entropy, even in the absence of a strangeness driven phase transition.
\item[Conclusions]  We conclude that the hyperon content of neutron stars and supernova matter
cannot be established with present constraints, and is essentially governed by the unconstrained 
coupling to the strange isoscalar meson. 
\end{description}
\end{abstract}


\maketitle


\section{Introduction}

\paragraph*{}
Understanding the physics underlying the QCD phase diagram remains a challenge. While at low chemical potentials lattice QCD has provided many answers \cite{lqcd}, at high densities, effective models \cite{Original RMF,Boguta,Walecka}
have been the source of most accomplishments. In order to rule out unreasonable propositions, experimental (at low densities) and observational (at high densities) constraints have been used as in
\cite{mariana1,mariana2,steiner,radii,haensel,ozel}. Two of the most reliable constraints at high
densities are the two-solar masses neutron stars (NS) detected in this
decade, PSR J1614-2230~\cite{Demo} and PSR
J0348+0432~\cite{Antoniadis}. Before them, effective models which
provided maximum stellar masses larger than 1.44 $M_\odot$ were
acceptable \cite{GlennBook,Glenn00,Glenn01}, but this is no longer the
reality and many revisions in the theory underlying neutron star
physics have been performed lately.

\paragraph*{}
The {\it hyperon puzzle} has then started to play a central role in the understanding of neutron star composition: while hyperons are energetically favoured, they soften the equation of state (EOS) and the existence of massive NS requires stiff EOSs.
This implies that the hyperon-nucleon and nucleon-nucleon interactions must be strongly repulsive at high hyperonic densities.
Meanwhile, many hypernuclei have been synthesised and recent analyses
of double-$\Lambda$ hypernuclei suggests an attraction, though small, in the
in the $\Lambda-\Lambda$ channel \cite{Ahn13}.
To fulfill these different requirements, it has become evident that
one has to go beyond simple SU(6) symmetry arguments to fix hyperon couplings, and more elaborate models have to be considered\cite{RikovskaStone:2006ta,Bednarek:2011gd,Weissenborn11b,Weissenborn11c,Bonanno11,Oertel:2012qd,Colucci2013,Luiz,
Banik2014,vanDalen2014,Katayama2014,Oertel}.
 
In the framework of relativistic mean field models (RMF),
this is achieved by the inclusion of strange scalar and
vector mesons \cite{strange-mesons, Pal} 
mediating the nuclear force alongside the usual 
scalar-isoscalar $\sigma$ or $f_0(550)$,  vector-isoscalar $\omega(782)$ and 
vector-isovector $\rho(770)$ mesons. In the case of the inclusion of the strange mesons, the scalar one normally labelled
$\sigma^{\ast}$ or $f_0 (980)$  and the vector labelled $\phi (1020)$, coupling constants between them and the
baryons are also necessary and they are usually fixed with a certain
degree of arbitrariness.  We discuss this point in greater detail
when the formalism is introduced in the next section, but at this
point it is necessary to emphasise that fixing them in a less
hand-waving way as possible would be desirable.

\paragraph*{}
In a recent paper \cite{low-density} the possible existence of a
liquid-gas phase transition was investigated with the help of RMF
models in a neutron-$\Lambda$ and neutron-proton-$\Lambda$ (np-$\Lambda$)
matter, whose meson-baryon coupling constants were allowed to vary in a broad
space, but restricted to satisfy empirical constraints on the nucleon-$\Lambda$
     ($U^N_\Lambda$) and $\Lambda-\Lambda$ ($U^\Lambda_\Lambda$) potentials
derived from hypernuclear experimental data.

The first one was fixed to be $-28$ MeV at nuclear saturation
density and the second one was fixed as $-0.67$ MeV at one fifth of the
nuclear saturation density ($n_0/5$), a reasonable
proposition \cite{Glenn02,Cugnon2000,Vidana2001,Khan2015}  inspired by experimental results \cite{Aoki09,Ahn13,Gal2016,Takahashi}. 
As a consequence, only two couplings could be taken as free parameters.
Then, the instabilities in the systems
 were investigated according to these restrictions and spinodal
 curves were observed, showing a clear liquid-gas phase transition at
 low densities, which was confirmed by the instability analysis of an ab-initio calculation \cite{abinitio,Lonardoni14} where no free parameters can be tuned.

\paragraph*{} 
At supersaturation densities, the investigation of a possible 
transition from a strange-poor to a strange-rich matter is also of
great importance in the studies of compact star constituents.
In particular, it was shown in \cite{Fran02,Fran04,Fran05} that non-relativistic
Skyrme functionals can produce one-body potentials compatible with
hyper-nuclear data, fulfil the $2 M_\odot$ neutron star constraint,
and still present a complex phase diagram with first and second order
phase transitions. These instabilities persist at finite temperature,
with important potential consequences on the supernova collapse
dynamics and the neutrino transport.  The existence of a strangeness
driven phase transition is however clearly model dependent. 
Another recent work shed some light on the role played by the
hyperon-meson coupling constants (including strange mesons) in stellar
macroscopic properties and the relation between results that are
consistent with observational values and the hyperon-hyperon potentials \cite{Oertel}. 
In this work, four different RMF parameterisations were used and it was
concluded that high mass neutron stars and not too large radii
are possible to obtain from an EOS containing a substantial amount of
hyperons, but the results depend a lot on the not yet known
hyperon-hyperon channels. The onset order of hyperons were also shown
to once again depend drastically on the parameterisation and hyperon-meson
coupling choices. In this work  no such instability is present, unless
unrealistically high values for the attractive couplings are
imposed~\cite{Oertel}. This suggests that the observed instabilities
might be a pathology of Skyrme functionals. 

\paragraph*{}
In the present work, 
we extend the analysis of ref.\cite{Oertel} to a systematic study of
the parameter space, searching for possible instabilities that
could give rise to a strangeness driven phase transition. Towards this
search, the $\chi_{\sigma \Lambda}$ and $\chi_{\sigma^{\ast} \Lambda}$ are
left as free parameters. The other two couplings, $\chi_{\omega
  \Lambda}$ and  $\chi_{\phi \Lambda}$ are constrained by
the potentials mentioned above, in the same way as in
\cite{low-density}.  

Once a large region of couplings is spanned,
we revisit stellar matter in $\beta$-equilibrium and electrically
neutral at zero and finite temperature and recalculate protoneutron
and neutron stars macroscopic quantities
from equations of state (EOS) that include strange mesons and where the couplings between hyperons and
mesons are chosen as mentioned above.
We then confront  stellar matter properties results obtained from the
phenomenological potentials and compare with results previously
published in the literature. 

Concerning the instability issue,  the result is that the whole RMF $n-\Lambda$ unconstrained parameter space is thermodynamically stable. 
This  leads to the strong conclusion that, within the RMF formalism, the only possibility to have a strangeness phase transition is at the quark level. 
Unfortunately, the final word about (in)stability of hyperonic matter
cannot be said, because a very different result is obtained from the
instability analysis of the ab-initio auxiliary field diffusion Monte Carlo
(AFDMC) model, within the three-body couplings which were recently shown to be able to correctly fulfill the $2 M_\odot$ neutron star constraint.

\paragraph*{}
The paper is organised as follows: In Section \ref{Formalism} we
present the main equations used in this paper, in Section
\ref{results} the results for (in)stable high density matter 
within RMF and an ab-initio models are shown and discussed 
and then the implications of these results in stellar matter are considered.
In the last
section, we draw the main conclusions and make the final remarks.

\section{Formalism} \label{Formalism}

\paragraph*{}
We next present the EOS and the parameterisation used in the present
work. We start from the Lagrangian density that includes hadronic and
leptonic matter and reads:

\begin{eqnarray}
\mathcal{L} && =\sum_{j}\overline{\psi }_{j}\left[ \gamma^{\mu }\left( i\partial _{\mu }-g_{\omega j}\omega _{\mu }-g_{\phi j}\phi_{\mu } \right.\right.  
 \left.\left.-g_{\rho j}\vec{\tau}\cdot \vec{\rho}_{\mu }\right) -m_{j}^{\ast}\right] \psi _{j} \nonumber \\
&&+\frac{1}{2}\left( \partial _{\mu }\sigma \partial ^{\mu }\sigma-m_{\sigma }^{2}\sigma ^{2}\right) -\frac{1}{3}bM_{N}\left(g_{\sigma N}\sigma\right)^{3} -\frac{1}{4}c\left(g_{\sigma N}\sigma\right)^{4} \nonumber \\   
&&-\frac{1}{4} \Omega_{\mu \nu} \, \Omega^{\mu \nu}+ \frac{1}{2} m_{\omega}^2 \, \omega_\mu \omega^\mu 
-\frac{1}{4}\Phi _{\mu \nu }\Phi ^{\mu \nu   }+\frac{1}{2}m_{\phi}^{2}\phi ^{\mu }\phi _{\mu } \nonumber \\
&&+\frac{1}{2}\left( \partial _{\mu }\sigma^{\ast}\partial^{\mu }\sigma^{\ast}-m_{\sigma }^{2} \sigma^{\ast 2} \right) 
-\frac{1}{4} \vec R_{\mu \nu } \,\cdot\,  \vec R^{\mu \nu } 
+ \frac{1}{2} m_\rho^2 \, \vec \rho_{\mu} \,\cdot\, \vec \rho^{\, \mu}
   \nonumber \\
&& +\sum_l \overline{\psi}_{l}\left(i\gamma_{\mu}\partial^{\mu}-m_l\right)\psi_{l},
\label{lagrangian}
\end{eqnarray}
where $m_j^* = m_j - g_{\sigma j}\, \sigma- g_{ \sigma^{\ast} j}\,
\sigma^{\ast}$ is the baryon effective mass, $m_j$ is the bare mass of
the baryon $j$ and $m_l$ is the mass of the lepton $l$.
The terms $\Omega_{\mu\nu }=\partial_\mu \omega_\nu - \partial_\nu \omega_\mu$~, $\Phi_{\mu\nu }=\partial_\mu \phi_\nu - \partial_\nu \phi_\mu$ and
$\vec R_{\mu \nu } = \partial_\mu \vec \rho_\nu -\partial_\nu \vec \rho_\mu 
-g_{\rho j} \left({\vec \rho_\mu \,\times \, \vec \rho_\nu } \right)$ are the strength tensors, where the up arrow in the last term denotes the isospin vectorial space.
The coupling constants are $g_{ij}=\chi_{ij} g_{i N}$, with the mesons denoted by index $i = \sigma, \omega, \rho, \sigma^{\ast}, \phi$ and the baryons denoted by $j$. Note that $\chi_{ij}$ is a proportionality factor between $g_{ij}$ and the nucleon coupling constants $g_{i N}$, with $N=n,~p$. 
The couplings $b$ and $c$ are the weights of the nonlinear scalar terms. The sum over $j$ can be extended over all baryons of the octet
$\left(n~, p~, \Lambda~,\Sigma^-~,\Sigma^0~,\Sigma^+~,\Xi^-~,\Xi^0~
\right)$ and the sum over $l$ includes the lightest leptons ($e^-,
e^+,\mu^-,\mu^+$). Neutrinos are not treated in the present work.

\paragraph*{}
We can then describe matter within the framework of  Relativistic Mean
Field (RMF)  theory \cite{Original RMF,Walecka, Boguta,Kapusta01}. 
Applying the Euler-Lagrange equations to the Lagrangian density
(\ref{lagrangian}) and using the mean-field approximation 
\cite{Walecka}, 
($\sigma \to \langle \sigma \rangle = \sigma_{0}~ ;\;
\omega_\mu \to \langle \omega_\mu  \rangle = \delta_{\mu 0} \,\omega_0 ~;\; 
\vec \rho_\mu  \to \langle \vec \rho_\mu  \rangle = \delta_{\mu 0} \, \delta^{i 3} \rho_0^3  
\equiv \delta_{\mu 0} \, \delta^{i 3} \rho_{03} ~;\;  \sigma^{\ast} \to \langle \sigma^{\ast} \rangle = \sigma^{\ast}_{0}~ ;
\; \phi_\mu \to \langle \phi_\mu  \rangle = \delta_{\mu 0} \,\phi_0   $), we obtain the
following well known equations of motion for the meson fields:
\begin{equation*}
\left( g_{\sigma N}\sigma _{0}\right) =\Delta _{\sigma }\left( \sum_{j}\chi _{\sigma
j}\rho _{j}^{s}-bM_{n}\left( g_{\sigma N}\sigma _{0}\right) ^{2}-c\left(
g_{\sigma N}\sigma _{0}\right) ^{3}\right),
\end{equation*}%
\begin{equation*}
\left( g_{\omega N}\omega _{0}\right) =\Delta _{\omega }\sum_{j}\chi _{\omega j}n _{j},
\end{equation*}%
\begin{equation*}
\left( g_{\rho N}\rho _{0}\right) =\Delta _{\rho }\sum_{j}\tau _{3j}\chi _{\rho j}n _{j},
\end{equation*}%
\begin{equation*}
\left( g_{\sigma N}\sigma^{\ast}_{0}\right) =\Delta _{\sigma \sigma^{\ast} }\sum_{j}\chi _{\sigma^{\ast}j}\rho _{j}^{s},
\end{equation*}%
\begin{equation}
\left( g_{\omega N}\phi _{0}\right) =\Delta _{\omega \phi }\sum_{j}\chi _{\phi j}n _{j},
\end{equation}
where the following factors are defined:
$\Delta _{\sigma } =\left( \frac{g_{\sigma N}}{m_{\sigma }}\right) ^{2}$, 
$\Delta _{\omega } =\left( \frac{g_{\omega N}}{m_{\omega }}\right) ^{2}$,
$\Delta _{\rho } =\left( \frac{g_{\rho N}}{m_{\rho }}\right) ^{2}$,
$\Delta _{\sigma\sigma^{\ast}} =\left( \frac{g_{\sigma N}}{m_{\sigma^{\ast}}}\right) ^{2}$, 
$\Delta _{\omega \phi } =\left( \frac{g_{\omega N}}{m_{\phi }}\right) ^{2}$.
The scalar and baryon densities are given respectively by
\begin{equation}
\rho _{j}^{s}=\frac{\gamma }{\left(2\pi\right)^3}\int\frac{m_{j}^{\ast }}{\sqrt{p^{2}+m_{j}^{\ast2 }}}\left[f_{j+}+f_{j-}\right] d^{3}p
\end{equation}
and
\begin{equation}
n _{j}=\frac{\gamma }{\left(2\pi\right)^3}\int
\left[f_{j+}-f_{j-}\right] d^{3}p, \quad n_B=\sum_j n_j
\end{equation}
where $f_{j\pm}$ is the Fermi distribution function,  with the positive
sign standing for particles and the negative sign to anti-particles:
\begin{equation}
f_{j\pm}=\frac{1}{1+\exp\left[{\left(E^{\ast}_{j}\mp\mu^{\ast}_{j}\right)/T}\right]}\text{,}
\end{equation}
with $E^{\ast}_{j}=\sqrt{p^{2}+m_{j}^{\ast2 }}$ and the chemical
potential $\mu _{j}^{\ast}$ is
\begin{equation}
\mu _{j}^{\ast }=\mu _{j}-\chi _{\sigma j}\left( g_{\omega N}\omega _{0}\right) -\tau _{3j}\chi _{\rho j}\left( g_{\rho N}\rho _{0}\right) -\chi _{\omega j}\left( g_{\phi N}\phi _{0}\right)\text{.}
\end{equation}

\paragraph*{}
The energy density  and pressure of the baryons are given respectively
by
\begin{equation}
\varepsilon _{B}=\frac{\gamma }{\left(2\pi\right)^3}\sum\limits_{j}\int\sqrt{p^{2}+m _{j}^{\ast2 }}\left[f_{j+}+f_{j-}\right] d^{3}p
\label{baryon_energy_density}
\end{equation}
and
\begin{equation}
p _{B}=\frac{\gamma }{\left(2\pi\right)^3}\sum\limits_{j} \int\frac{p^{2}}{\sqrt{p^{2}+m_{j}^{\ast2 }}}\left[f_{j+}+f_{j-}\right] d^{3}p ,
\end{equation}
and for the leptons the expressions can be read from the ones above
once the proper substitutions are made for the masses ($m_l$ instead
of $m^{\ast}_j$) and chemical potentials ($\mu_l$ instead of $\mu^{\ast}_j$). For the mesons
we have:
\begin{eqnarray}
\varepsilon _{M}&=&\frac{\left( g_{\sigma N}\sigma _{0}\right) ^{2}}{%
2\Delta _{\sigma }}+\frac{\left( g_{\omega N}\omega
_{0}\right) ^{2}}{2\Delta _{\omega }}+\frac{\left( g_{\rho N}\rho
_{0}\right) ^{2}}{2\Delta _{\rho }} \nonumber \\ 
&&+\frac{\left( g_{\sigma N}\sigma^{\ast}_{0}\right)
^{2}}{2\Delta _{\sigma \sigma^{\ast}}}+\frac{\left( g_{\omega N}\phi
_{0}\right) ^{2}}{2\Delta _{\omega \phi }} \nonumber \\
&&+\frac{1}{3}bM_{n}\left( g_{\sigma N}\sigma _{0}\right) ^{3}+\frac{1}{4}c\left( g_{\sigma N}\sigma _{0}\right)^{4},
\label{meson_energy_density}
\end{eqnarray}
and
\begin{eqnarray}
p _{M}&=&-\frac{\left( g_{\sigma N}\sigma _{0}\right) ^{2}}{%
2\Delta _{\sigma }}+\frac{\left( g_{\omega N}\omega
_{0}\right) ^{2}}{2\Delta _{\omega }}+\frac{\left( g_{\rho N}\rho
_{0}\right) ^{2}}{2\Delta _{\rho }} \nonumber \\
&&-\frac{\left( g_{\sigma N}\sigma^{\ast}_{0}\right)
^{2}}{2\Delta _{\sigma \sigma^{\ast}}}+\frac{\left( g_{\omega N}\phi
_{0}\right) ^{2}}{2\Delta _{\omega \phi }} \nonumber\\
&&-\frac{1}{3}bM_{n}\left( g_{\sigma N}\sigma _{0}\right) ^{3}-\frac{1}{4}c\left( g_{\sigma N}\sigma _{0}\right)^{4}.
\end{eqnarray}
Finally the expressions for the EOS including interacting baryons and
non-interacting leptons read:
\begin{equation*}
\varepsilon =\varepsilon _{B}+\varepsilon _{M} + \varepsilon _{l},
\qquad
p =p _{B}+p _{M} + p_{l}
\end{equation*}
and the conditions of charge neutrality 
\begin{equation}
\mu_p = \mu_n - \mu_e = \mu_{\Sigma^+}, 
\mu_{\Sigma^-} = \mu_{\Xi^-} =\mu_n + \mu_e, 
\mu_n=\mu_\Lambda= \mu_{\Sigma^0} = \mu_{\Xi^0},
\end{equation}
and $\beta$-equilibrium
\begin{equation}
\sum_ j q_j n_j + \sum_l q_l n_l =0
\end{equation}
 are incorporated.

The expressions for zero temperature systems can be obtained from the
above equations by substituting the particle Fermi distribution functions by a
step function in Fermi momentum. The antiparticles are then neglected.

\paragraph*{}
Now we go back to the discussion of how to choose the baryon-meson
coupling constants.
The coupling constants of the nucleons with mesons $\sigma$, $\omega$ and $\rho$ are fixed
in such a way that the bulk properties of nuclear matter are obtained
and in the present work we have opted to use the parametrisation known
as GM1 \cite{Moszkowski}. The hyperon couplings have been chosen in different ways in the literature, either 
based on simple symmetry considerations \cite{Sym00,Moszkowski,Pal,Banik2014,Sym02,Luiz,Colucci2013},
or requiring an EOS in $\beta$-equilibrium sufficiently stiff to
justify the observation of very massive 
neutron stars \cite{Glenn02,Glenn03}. 
In the case of the inclusion of the strange mesons, $\sigma^{\ast}$
and $\phi$, the same degree of arbitrariness is found, i.e., their
couplings with the hyperons have generally obeyed  SU(6) symmetry
based on quark counting \cite{strange-mesons, Pal}. A choice of couplings
based on a mixture of arguments \cite{Cavagnoli2} and on SU(3)
symmetry \cite{Weissenborn11b,Weissenborn11c, Luiz} has also been done. If massive
stars are to be obtained, the SU(6) symmetry has to be partially
broken and, in this case, strange mesons can couple with the nucleons,
influencing nuclear bulk properties and forcing a reparameterisation
\cite{Luiz}. In the present work we avoid the discussion based on
group theory and focus our choice on the phenomenological information 
that can be extracted on the $U_{\Lambda}^{\Lambda}$ potential  for the meson-$\Lambda$ couplings.
When the other hyperons are included, we make different choices for
their couplings and we discuss this point when the results are presented.
We then choose the hyperon-meson (strange and non-strange) coupling
constants as in \cite{low-density}. In analogy with what has
been done with the $\chi_{\sigma \Lambda}$ when constrained by the 
empirical information on the hyper-nuclear potential $U_{\Lambda}^{N}$
\cite{Glenn02,Cugnon2000} that can be extracted from hyper-nuclear
data, we have constrained the strange meson couplings 
to the  binding energy data of double $\Lambda$ hypernuclei
available in the literature \cite{Aoki09,Ahn13}. 
Considering the general form of the $\Lambda$-potential $U_{\Lambda}$
in the RMF models:
\begin{eqnarray}
U_{\Lambda}\left(n_{N},n_{\Lambda}\right)&=&\chi_{\omega \Lambda}\left(g_{\omega N}\omega_{0}\right)+\chi _{\phi \Lambda}\left(g_{\omega N}\phi_{0}\right) \nonumber \\
&&-\chi_{\sigma \Lambda}\left(g_{\sigma N}\sigma_{0}\right)-\chi_{\sigma^{\ast} \Lambda}\left(g_{\sigma N}\sigma_{0}^{\ast}\right),
\label{UL_fileds}
\end{eqnarray}

the above mentioned phenomenological information can be written as
\cite{Cugnon2000,Vidana2001,Khan2015}: 
\begin{eqnarray}
U_{\Lambda}^{N}\left(n_{0}\right)&\equiv& 
U_{\Lambda}\left(n_{N}=n_0,n_{\Lambda}=0\right)
\approx -28~\text{MeV} \label{constraints1}\\
U_{\Lambda}^{\Lambda}\left(\frac{n_{0}}{5}\right)&\equiv& 
U_{\Lambda}\left(n_{N}=0,n_{\Lambda}=\frac{n_0}{5}\right)
\approx -0.67~\text{MeV}.
\label{constraints2}
\end{eqnarray}

This leads to the following conditions on the strange couplings:
\begin{eqnarray}
\chi_{\omega \Lambda}&=&\frac{\chi_{\sigma \Lambda}\left.\sigma\right|_{n_N=n_{0}}+
U_{\Lambda}^{N}\left(n_{0}\right)}{\left.\omega\right|_{n_N=n_{0}}}.
\label{Uln_contraint_fields}  \\
\chi_{\phi \Lambda}&=&\left( \frac{m_{\phi }}{m_{\omega }}\right) \nonumber\\
&&\times\sqrt{\frac{U_{\Lambda}^{\Lambda}\left(n_{\Lambda 0}\right)+\alpha \chi_{\sigma \Lambda}\left. \Sigma \right\vert _{n_{\Lambda }=%
n_{\Lambda 0}}-\chi_{\omega \Lambda}\left. \omega \right\vert _{n_{\Lambda }=
n_{\Lambda 0}%
}}{\chi_{\omega \Lambda} \left. \omega \right\vert
_{n_{\Lambda }=n_{\Lambda 0}}}}\chi_{\omega \Lambda}. \nonumber \\
\label{Ull_contraint_fields}
\end{eqnarray}
where we have defined $n_{\Lambda 0 }=n_0/5$, $\alpha=1+\left(\frac{\chi_{\sigma^{\ast} \Lambda}}{\chi_{\sigma \Lambda}}\right) ^{2}\left( \frac{m_{\sigma }}{m_{\sigma ^{\ast }}}\right)^{2}$, and 
\begin{equation}
\Sigma=  
\sigma -\left(\frac{g_{\sigma N}}{m_{\sigma }}\right) ^{2}\left( \frac{\alpha-1}{\alpha }\right) \left(-bm_{n}\sigma ^{2}-c\sigma ^{3}\right).
\end{equation}

It is worth mentioning that  the values for the potentials in
  Eqs.(\ref{constraints1}),(\ref{constraints2}) are actually 
indirectly extracted from the experimentally measured total binding
 and single particle energies. The quoted 
potentials are extracted from theoretical calculations of 
hypernuclei adjusted such as to reproduce the experimental data.
 Notice   also that the value $B_{\Lambda\Lambda}=-0.67$ MeV
\cite{Ahn13} for the measured bond energy leads to a
potential with low attractiveness, contrary to another choice commonly
used in the literature that foresees this potential as $-5$ MeV also at
$n_0/5$ \cite{Franklin}.  According to a
  recent review \cite{Gal2016}, the unambiguous observation of
  $^6_{\Lambda \Lambda} He$ \cite{Takahashi} changed the previously
  accepted value of $B_{\Lambda\Lambda}$ to the values we are using in
  the present paper.  Nevertheless, we have checked that our results do not
qualitatively change if a modification of $\pm 5$ MeV is applied to
the assumed $\Lambda$-potential constraints.

\paragraph*{}
The $\chi_{\sigma \Lambda}$ and  $\chi_{\sigma^{\ast} \Lambda}$ are left to be
free parameters in the model, which we can vary such as to
explore the possible appearance of instabilities. 

\paragraph*{}
To investigate the existence of instabilities, we rely on the
curvature matrix and related spinodal section.
A first order phase transition is signaled by an instability or concavity anomaly in the mean-field thermodynamic total energy density. The total energy density of a three-component system is a three variable function of the densities. If the scalar function $\epsilon$ is smooth, or at least twice continuously differentiable, we can introduce the curvature matrix $\mathbf{C}$ associated to  $\epsilon$ at a point denoted by $P\in\left(n_{B}\times n_{S}\times n_{L}\right)$:
\begin{equation}
C_{ij}=\left( \frac{\partial ^{2}\epsilon \left( n_{i},n{j}\right) }{\partial n_{i}\partial n_{j}}\right)
\text{,} \label{eq:spinodal}
\end{equation}
where $i,j=B,S,L$, standing for baryon, strange and lepton contributions.

Since the curvature $\mathbf{C}$ is real and symmetric, its normalised eigenvectors $\mathbf{u}_k$ are orthogonal and define a basis in the observable density space. The associated eigenvalues 
$\lambda_k$ represent the energy curvature associated to a density variation in the direction of 
$\mathbf{u}_k$. Let us note $\lambda_<$ the minimum eigenvalue at a
given point $(n_B, n_S, n_L)$, and $\mathbf{u}_<=(\delta n_B^<,\delta
n_S^<, \delta n_L^<)$ the associated eigenvector. If $\lambda_< <0$,
then any density fluctuation becomes spontaneously amplified in the
direction of the unstable eigenvalue  $\mathbf{u}_<$, that is the
system spontaneously develops an inhomogeneity between two regions
characterised by the particle composition $(n_B \pm \delta n_B^<,n_S
\pm \delta n_S^<, n_L \pm \delta n_L^<)$. 
The inhomogeneity then amplifies in time according to the mechanism of
spinodal decomposition \cite{Chomaz04}, until two stable points with 
all positive eigenvalues (a dense and a diluted phase) are reached.  
At zero temperature the dilute phase typically corresponds to the
vanishing of one or more partial densities \cite{Ducoin06} and
the instability direction can be approximately identified by
the direction of phase separation \cite{Chomaz04}.  
As a consequence, 
at any density point inside the spinodal, the density and composition of the two coexisting phases issued from the phase transition can be approximately inferred from the intersection of the unstable eigenvector direction and the spinodal contour, defined by the condition $\lambda_<=0$.
If the spinodal region is unbound at high density, the instability
cannot be cured by phase separation. In that case the system violates
the basic convexity principle of thermodynamics, and the instability
has to be considered as a pathology.
For a better understanding of the physics underlying strange matter
and a comparison with the results obtained at low densities \cite{low-density},
we next restrict our study to a simple model containing only neutrons
and $\Lambda$ hyperons. In this simplified case, the curvature matrix is
reduced to a $2x2$ 
matrix which can be equivalently expressed in the good quantum number 
plane $n_B,n_S$ or in the rotated frame corresponding to the particle densities $n_n,n_\Lambda$. 

\paragraph*{}
In the next Section, we investigate the possible existence of
instabilities at high densities and compare our results with  the ab-initio
AFDMC model. Then, we analyse stellar matter restricted to the above
mentioned choice of coupling constants.

\section{Results and Discussion} \label{results}

\paragraph*{}
We first discuss a hypothetical matter, where the only allowed
hyperon is the $\Lambda$. 
Indeed, the other hyperon potentials, 
$U_{\Sigma}^{\Sigma}(n_0))$ and $U_{\Xi}^{\Xi}(n_0)$ cannot be
obtained from experiments with a reasonable degree of certainty and
their inclusion adds arbitrariness to the results. Hence, we study
n$\Lambda$ and np$\Lambda$ matter first at zero temperature to understand the main consequences of our
choices for the coupling constants and only afterwards we incorporate
the other hyperons and finite temperature effects through different
snapshots of the stellar evolution. 

Next, the particle fraction is
defined as 
$$Y_p=n_p/\sum_j n_j, \quad p=j,l . $$

\subsection{n$\Lambda$ and np$\Lambda$ matter}

\subsubsection{Instability analysis and phase diagram}

\begin{figure*}
\centering
\subfloat[]
{\includegraphics[width=0.33\textwidth]{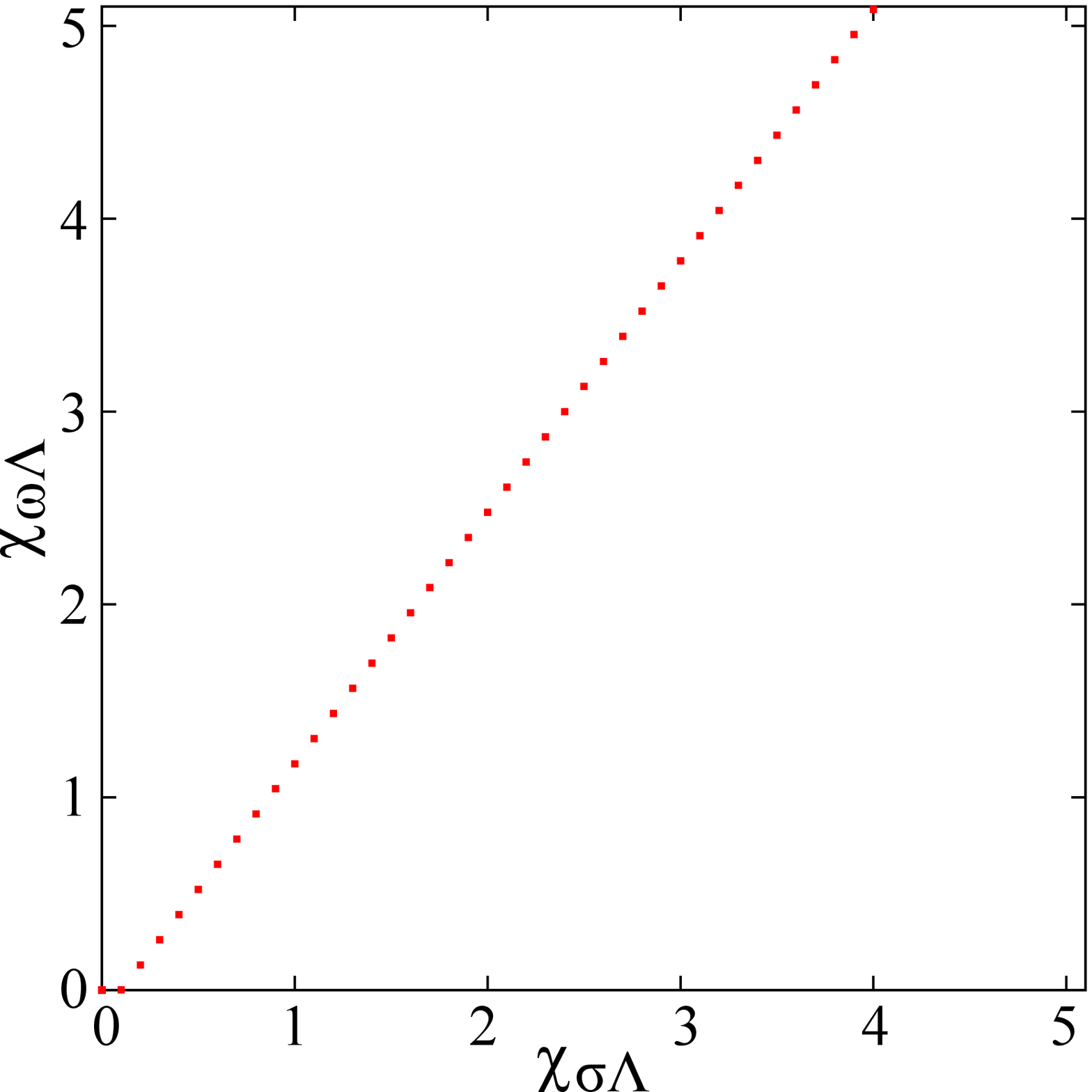}}
\subfloat[]
{\includegraphics[width=0.33\textwidth]{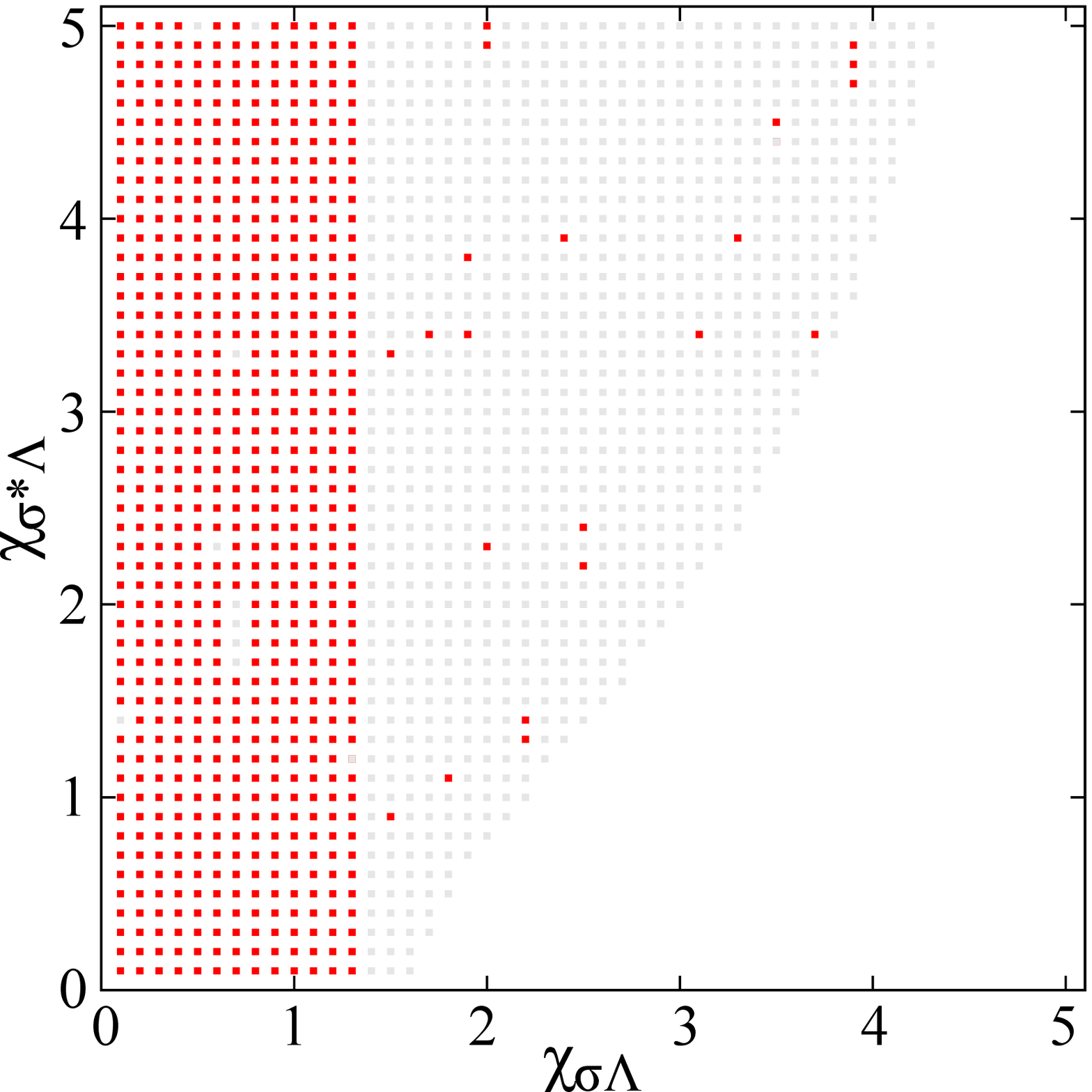}}
\subfloat[]
{\includegraphics[width=0.33\textwidth]{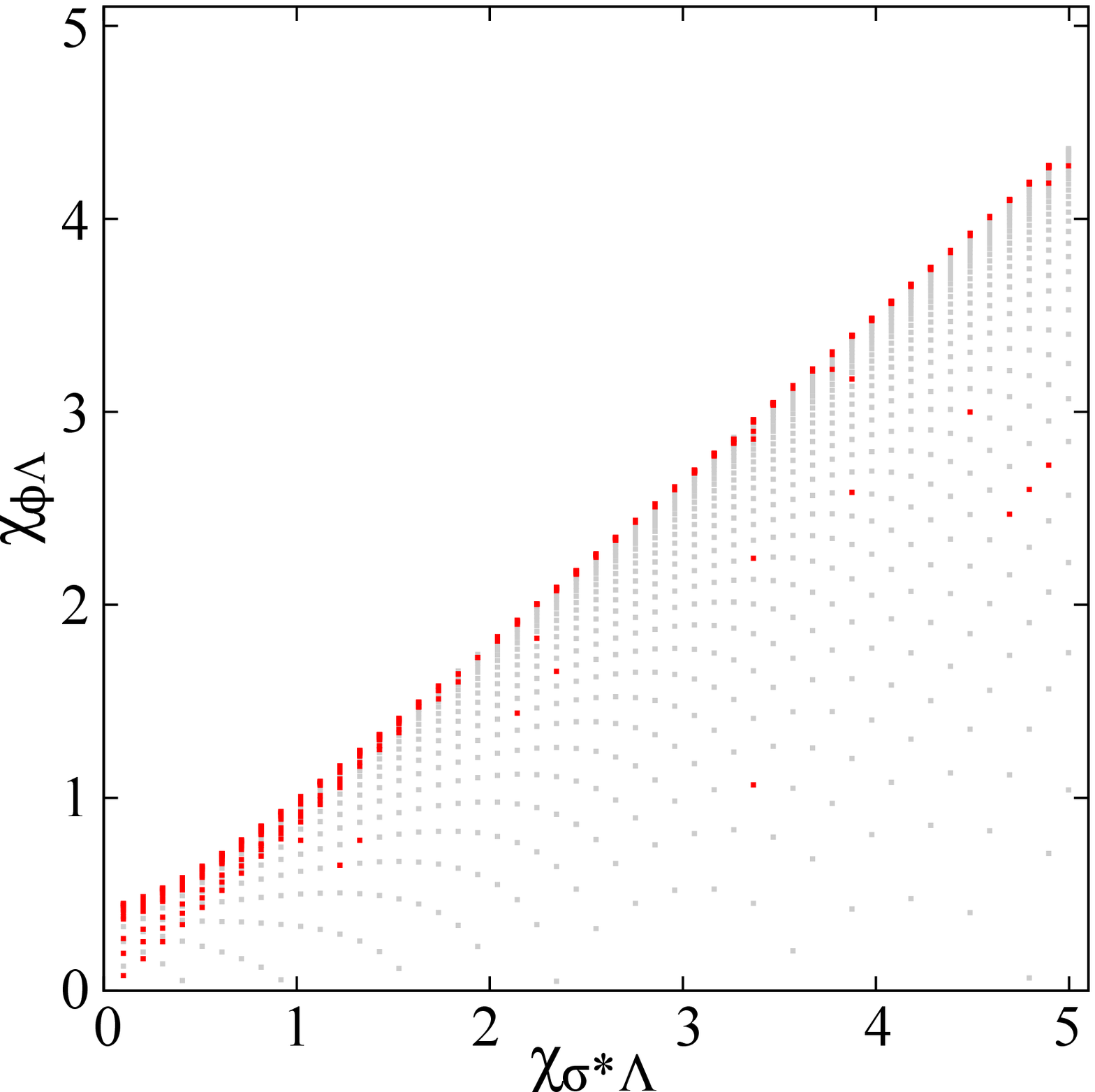}}
\caption{(Color online) Left panel: relation between $\chi_{\sigma \Lambda}$ and $\chi_{\omega \Lambda}$ in the RMF under the constraint $U_{N}^{\Lambda}=-28$ MeV. Central panel: parameter space for the $\Lambda$ effective mass. Gray points indicate the employed parameter grid; red points indicate the parameter sets leading 
to a positive $\Lambda$ effective mass in neutron star matter. Right panel: parameter space for the strange meson couplings. The meaning of the color code is as in the central panel.}
\label{Grid}
\end{figure*}

\paragraph*{}
We start by showing the relations between the couplings in Figure \ref{Grid}, where
$\chi_{\sigma \Lambda}$ and  $\chi_{\sigma^{\ast} \Lambda}$ are left to be free parameters in the model, which we have largely varied such as to explore the possible appearance of instabilities.
The left part of Fig.\ref{Grid} shows the relation between $\chi_{\sigma \Lambda}$ and $\chi_{\omega \Lambda}$ when we consider 
Eq.(\ref{Uln_contraint_fields}).  Similarly, the right part of Fig.\ref{Grid} shows the relation between $\chi_{\sigma^{\ast} \Lambda}$ and $\chi_{\phi \Lambda}$ when we consider 
Eq.(\ref{Ull_contraint_fields}). As already observed in \cite{low-density},
the linear dependence obtained means that in the framework of the NLWM a 
strong attraction at low densities is always correlated to a strong
repulsion at high densities.  It is interesting to remark that the same is true in non-relativistic 
models \cite{Fran02,Fran04,Fran05}.

\paragraph*{}
The residual parameter space is shown in the central part of Fig.\ref{Grid}. The 
$\chi_{\sigma \Lambda}$ range can be further limited requiring that
convergent solutions are obtained in 
 hyperonic stellar matter (with all the baryon octet ($H$), electrons, muons and $\rho$ meson
included) in $\beta$ equilibrium. For this calculation,  the $\rho$ meson coupling
is fixed at $\chi_{\rho H}=1.5$  for $H=\Sigma,\Xi$, so that we guarantee that
$\Lambda$'s are the first hyperons to appear
and the (unconstrained) couplings to $\Sigma$ do not play a major role.
The gray points in Fig.\ref{Grid} are related to divergent solutions, where
the $\Lambda$ effective mass goes to zero at some finite density.  
The red points yield possible solutions and, in some cases, 
the maximum masses can reach two solar masses with a finite
$Y_{\Lambda}$ \cite{Luiz,Gusakov}.

\paragraph*{}
To obtain some preliminar stellar macroscopic properties, the EOS with
the constrained couplings are used as input to the Tolman-Oppenheimer-Volkoff
(TOV) \cite{tov} equations. The high density behaviour of the
different models in terms of 
maximum mass is summarised in Table \ref{Table Mmax/M}, where 
$\epsilon_{0}$ stands for the central energy density and
$M_{\text{max}}$for the maximum star mass.
We can see that all couplings $\chi_{\sigma \Lambda}\ge 0.7$ lead to realistic values of the maximum mass, and a non-negligible hyperon content is present if $\chi_{\sigma \Lambda}< 1$. It is very interesting to observe that if 
the scalar coupling is sufficiently strong $\chi_{\sigma \Lambda}\geq 1$, hyperons can be completely neglected whatever the coupling assumed to the strange mesons.

\begin{table*}
\begin{ruledtabular}
\begin{tabular}{ccccccccc}
&$\chi_{\sigma \Lambda}$ & $\chi_{\omega \Lambda}$ & $\chi_{\sigma^{\ast} \Lambda}$ & $\chi_{\phi \Lambda}$&$Y_{\Lambda}$ 
& $\epsilon_{0}~(\texttt{fm}^{-4})$ & $M_{\texttt{max}} /M_{\odot}$ &\\
\hline

&0.50&0.52&0.50&0.61&0.63&5.23&1.69&\\
&0.50&0.52&1.50&1.37&0.63&5.23&1.69&\\
&0.50&0.52&3.50&3.07&0.64&5.23&1.68&\\
&0.70&0.78&0.50&0.65&0.52&5.80&2.08&\\
&0.70&0.78&1.00&0.99&0.55&5.92&2.08&\\
&0.70&0.78&3.50&3.08&0.54&5.82&2.08&\\
&1.00&1.17&0.50&0.67&0.03&5.85&2.35&\\
&1.00&1.17&1.50&1.40&0.03&5.81&2.36&\\
&1.00&1.17&3.50&3.08&0.03&5.79&2.35&\\
&1.30&1.56&0.50&0.62&0.00&5.83&2.36&\\
&1.30&1.56&1.50&1.37&0.00&5.83&2.36&\\
&1.30&1.56&3.50&3.07&0.00&5.83&2.36
\end{tabular}
\vspace{0.5cm}
\end{ruledtabular}
\caption{Chosen parameter sets for the $\Lambda$ coupling constants and corresponding central $\Lambda$ fraction $Y_{\Lambda}$ and central energy density $\epsilon_{0}$
for a neutron star at its maximum mass. In our calculations we keep
$\chi_{\rho}=1.5$ fixed.}
\label{Table Mmax/M}
\end{table*}
 
\begin{figure}[h!]
\centering
{\includegraphics[width=0.45\textwidth]{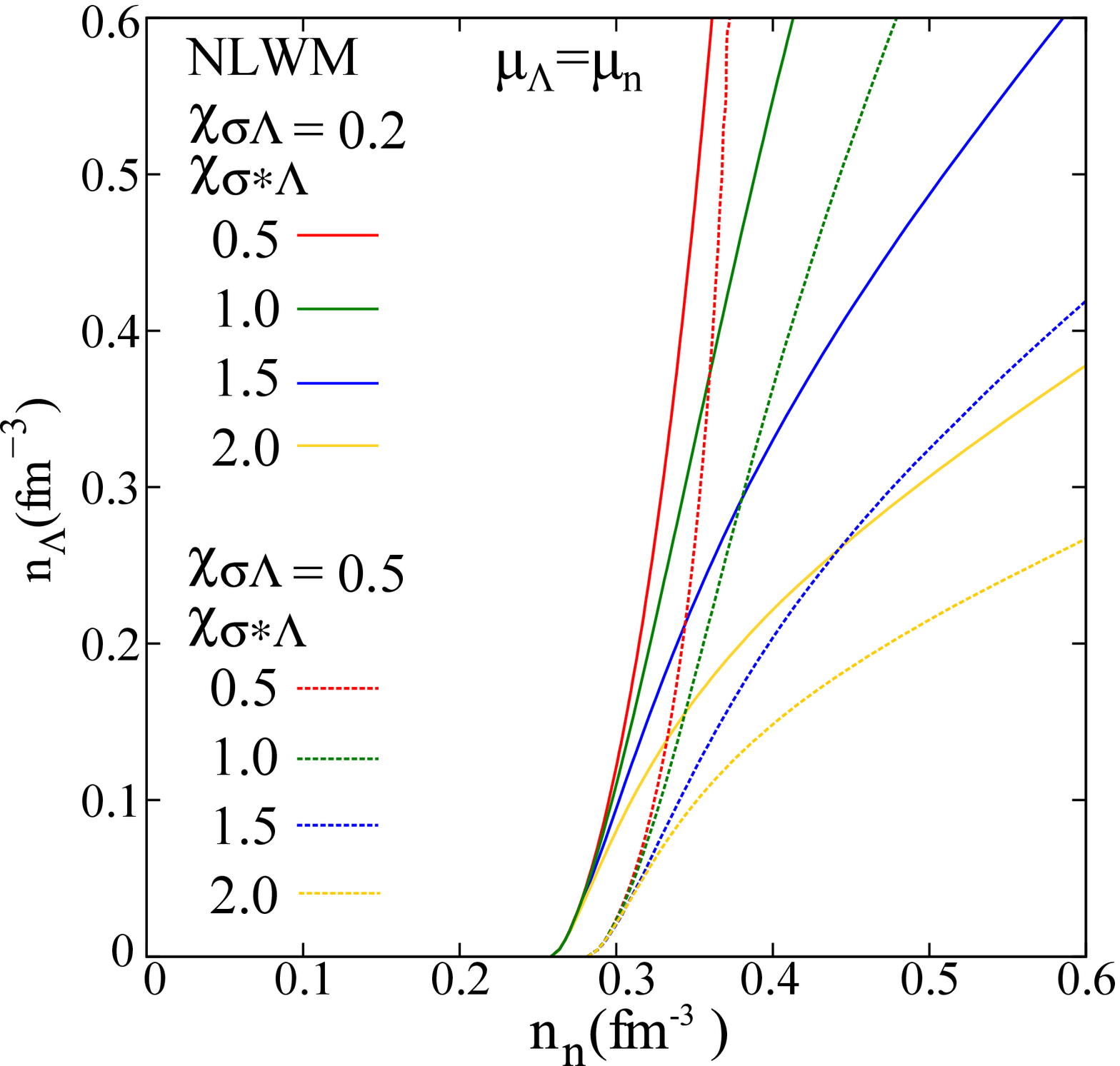}}
\caption{(Color online) Equilibrium trajectories $\mu_{n}=\mu_{\Lambda}$ for
  different values of the coupling constants.}
\label{Figure1}
\end{figure}

\paragraph*{}
The equilibrium trajectories for which $\mu_{n}=\mu_{\Lambda}$ 
are displayed in Figure \ref{Figure1} with
different values of the coupling constants.   
One can clearly see that the onset of
$\Lambda$s is determined by the $\chi_{\sigma \Lambda}$ because the
larger its value, the larger the repulsion felt by the $\Lambda$s and 
consequently, the latter in density is their onset. 
This result is due to the linear correlation between 
$\chi_{\sigma \Lambda}$ and $\chi_{\omega \Lambda}$ observed in Figure
\ref{Grid}, which is a consequence of the constraints imposed in 
equations (\ref{constraints2}).
On the other hand,
the $\chi_{\sigma^{\ast}  \Lambda}$ influences the amount of
$\Lambda$ hyperons in the system without modifying their onset.

\paragraph*{}
The curvature matrix Eq.(\ref{eq:spinodal}) was diagonalised in the interval $n_0\leq n_n \leq 1$ fm$^{-3}$,  $0\leq n_\Lambda \leq 1$ fm$^{-3}$ for all the different points in the parameter space noted as red points in Fig.\ref{Grid}. No negative eigenvalue was found. This means that,
even adding the strange mesons, if the physical constraints are applied, the system is always stable.
We can conclude that the possible presence of instability is not a
side effect of strongly attractive hyperonic
interactions in the dense medium.
 
\begin{figure}[h]
\begin{tabular}{c}
\subfloat[]
{\includegraphics[width=0.4\textwidth]{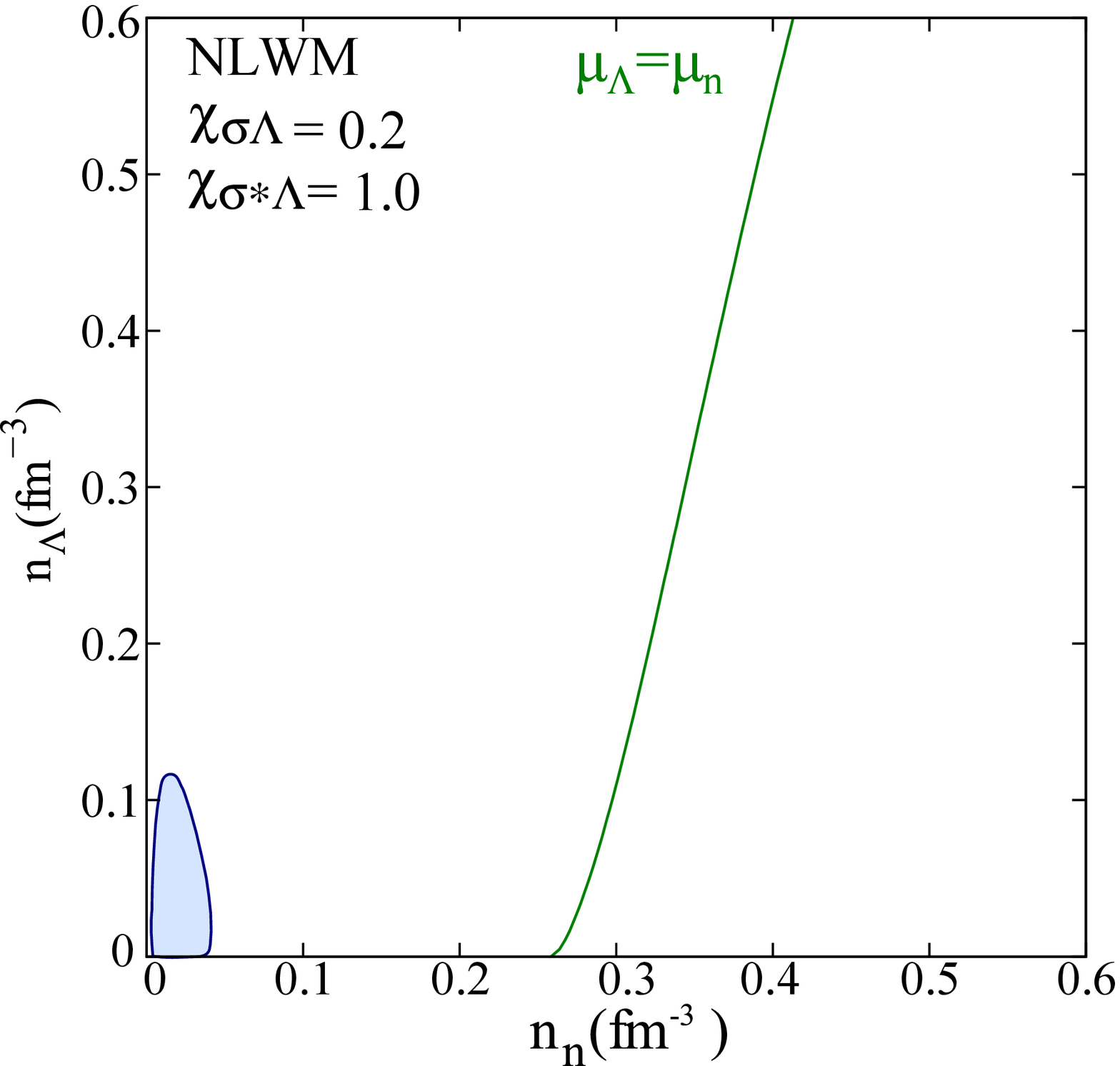}}\\
\subfloat[]
{\includegraphics[width=0.4\textwidth]{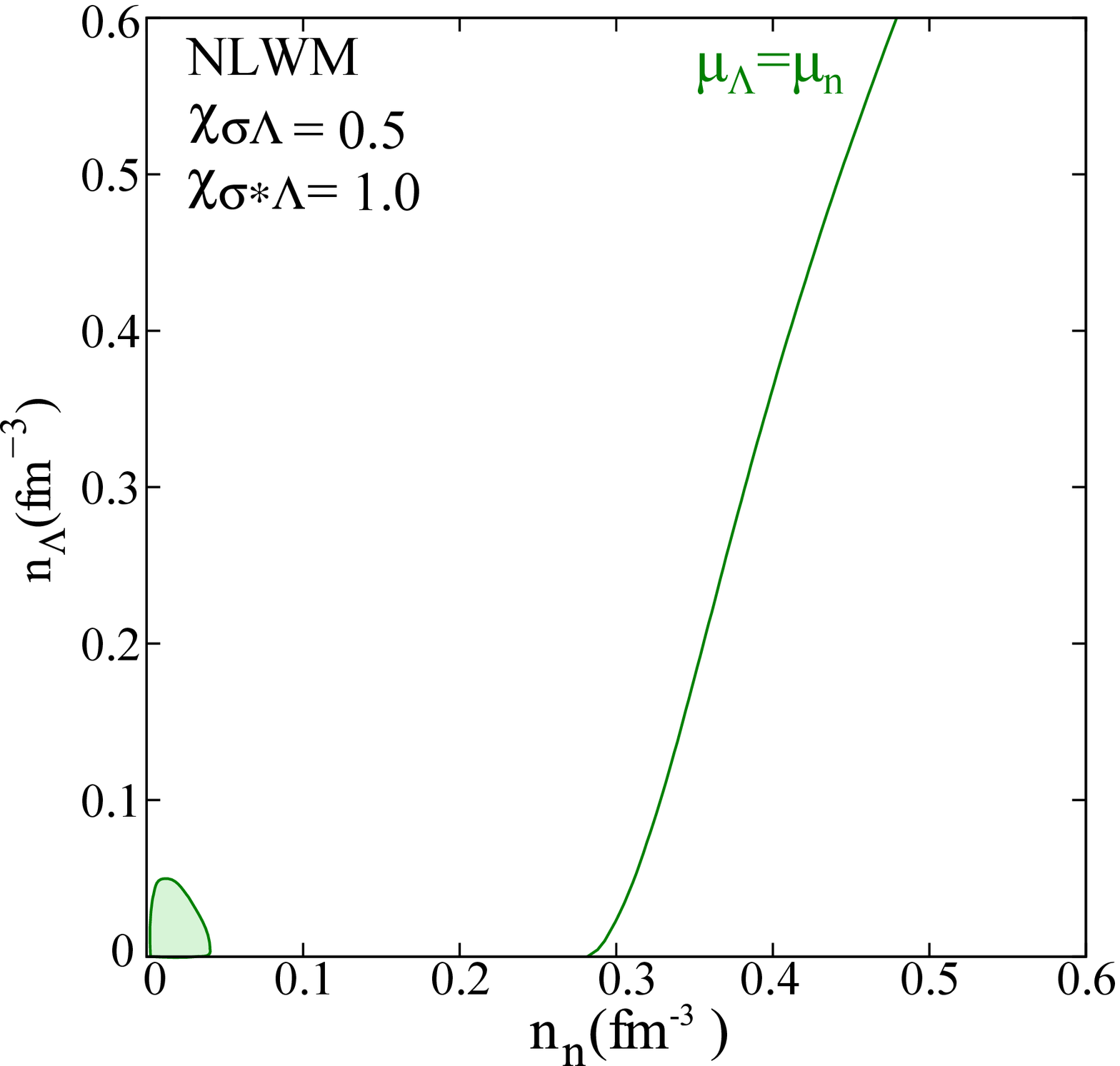}}
\end{tabular}
\caption{(Color online) Spinodals on the density plane alongside the equilibrium
  trajectories for two values of the coupling constants.}
\label{Figure2}
\end{figure}

In Figure \ref{Figure2} we plot the spinodals on the same plane as the
equilibrium trajectories for different values of the coupling
constants and one can see that they never cross. 
This fact was already
expected because there are no instability regions at high densities
and the onset of $\Lambda$ particles happens around twice nuclear saturation
density.

\subsubsection{Comparison to the 
  AFDMC model} \label{sec:abinitio}

\paragraph*{}
 
In recent years, ab-initio models based on different quantum Monte Carlo simulation techniques \cite{abinitio,Gandolfi09,Gezerlis10,Gandolfi12,Gandolfi14, Lonardoni14}
have been applied to (hyper)-nuclear matter.  In the pure neutron
sector,  such models provide an essential constraint to
phenomenological mean field models, which starts to 
be routinely applied in order to fix some of the unknown couplings.
Some calculations including $\Lambda$-hyperons are also available.
In particular, a recent AFDMC calculation has been used to compute hyper-nuclear observables \cite{Lonardoni14} and allows producing very massive neutron stars in agreement with the observations  \cite{abinitio},  though with the inclusion of a very stiff three-body term with leads to a negligible strangeness fraction.
This model is based on a phenomenological bare interaction inspired
by the Argonne-Urbana forces \cite{Usmani08}, with the addition of a purely phenomenological three-body force. 

\paragraph*{}
In a recent paper \cite{low-density}, we have shown that instabilities
exist within this model at sub-saturation density, corresponding to
the nuclear liquid-gas phase transition. 
The similarity of the obtained phase diagram with the one of the RMF
was interpreted as an indication of reliability of our phenomenological model.
It is therefore interesting to see if the same agreement is observed
concerning the phase diagram at high density.
The energy density of a neutron-$\Lambda$ mixture was fitted in \cite{abinitio}  by the following simple functional form:
\begin{eqnarray}
\epsilon _{AFDMC}\left( n_{n},n_{\Lambda }\right)  &=&\left[ a\left(\frac{n_{n}}{n_{0}}\right) ^{\alpha }+b\left( \frac{n_{n}}{n_{0}}\right)^{\beta }\right] n_{n} \nonumber \\
&&+\frac{1}{2m_{\Lambda }}\frac{3}{5}n_{\Lambda }\left(3\pi ^{2}n_{\Lambda }\right)^{2/3}\nonumber \\
&&+\left( m_{n}n_{n}+m_{\Lambda }n_{\Lambda }\right) \nonumber \\
&&+c_{1}^{^{\prime}}n_{\Lambda }n_{n}+c_{2}^{^{\prime }}n_{\Lambda }n_{n}^{2}.
\label{pederiva}
\end{eqnarray}%
In this expression, the first term represents the energy density of pure neutron matter, where $n_{0}$ is saturation point of symmetric nuclear matter, and the parameters $a$, $\alpha$, $b$ and $\beta$ are given in \cite{abinitio}.
The second term highlights the kinetic energy density of pure $\Lambda$-matter, and the last two terms,
obtained from the fitting of the Monte Carlo results for different $Y_{\Lambda}
=n_\Lambda/n_B$ fractions, provide an
analytical parametrisation for the difference between Monte Carlo energies of pure $\Lambda$ and pure neutron matter.
Notice that $\Lambda$-$\Lambda$ interactions are neglected in \cite{abinitio}, which explains why pure $\Lambda$ matter ($n_n=0$) 
behaves as a Fermi gas of noninteracting particles. This means that the extrapolations to high $\Lambda$ densities have to be considered with a critical eye.
Still this choice corresponds, in the RMF language, to 
$\chi_{\sigma^{\ast} \Lambda}=\chi_{\phi \Lambda}=0$, which is part of our parameter space. Moreover, a part of the $\Lambda$-$\Lambda$ interaction might be effectively included in the three-body term.

The numerical values of the constants $c_{1}^{^{\prime }} $ and
$c_{2}^{^{\prime}}$ are given in \cite{abinitio}.
Using Eq.(\ref{pederiva}), the chemical potentials become:
\begin{eqnarray}
\mu _{n}\left( n _{n}\right) &=&a\left( \alpha +1\right) \left( \frac{n_{n}}{n_{0}}\right) ^{\alpha }+b\left( \beta +1\right) \left( \frac{n_{n}}{n_{0}}\right) ^{\beta }\nonumber \\
&&+m_{n}+c_{1}^{^{\prime }}n_{\Lambda }+2c_{2}^{^{\prime
}}n_{\Lambda }n_{n},
\end{eqnarray}%
and
\begin{equation}
\mu _{\Lambda }\left( n_{\Lambda }\right) =\frac{1}{2m_{\Lambda }}\left(3\pi^{2}n_{\Lambda }\right) ^{2/3}+m_{\Lambda}+c_{1}^{^{\prime }}n_{n}+c_{2}^{^{\prime }}n_{n}^{2}.
\end{equation}%

\begin{figure*}
\centering
\subfloat[]
{\includegraphics[width=0.33\textwidth]{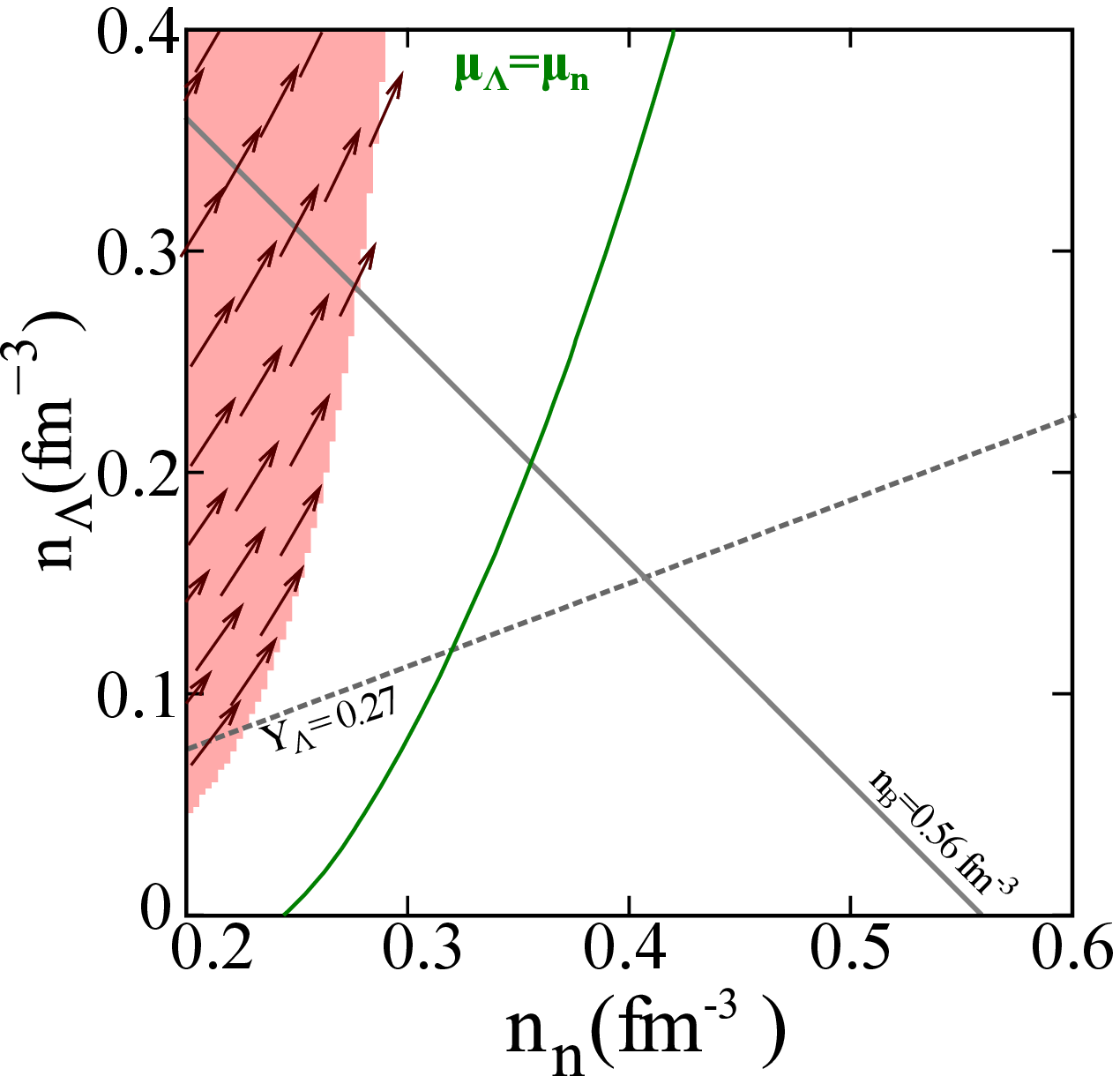}}
\subfloat[]
{\includegraphics[width=0.33\textwidth]{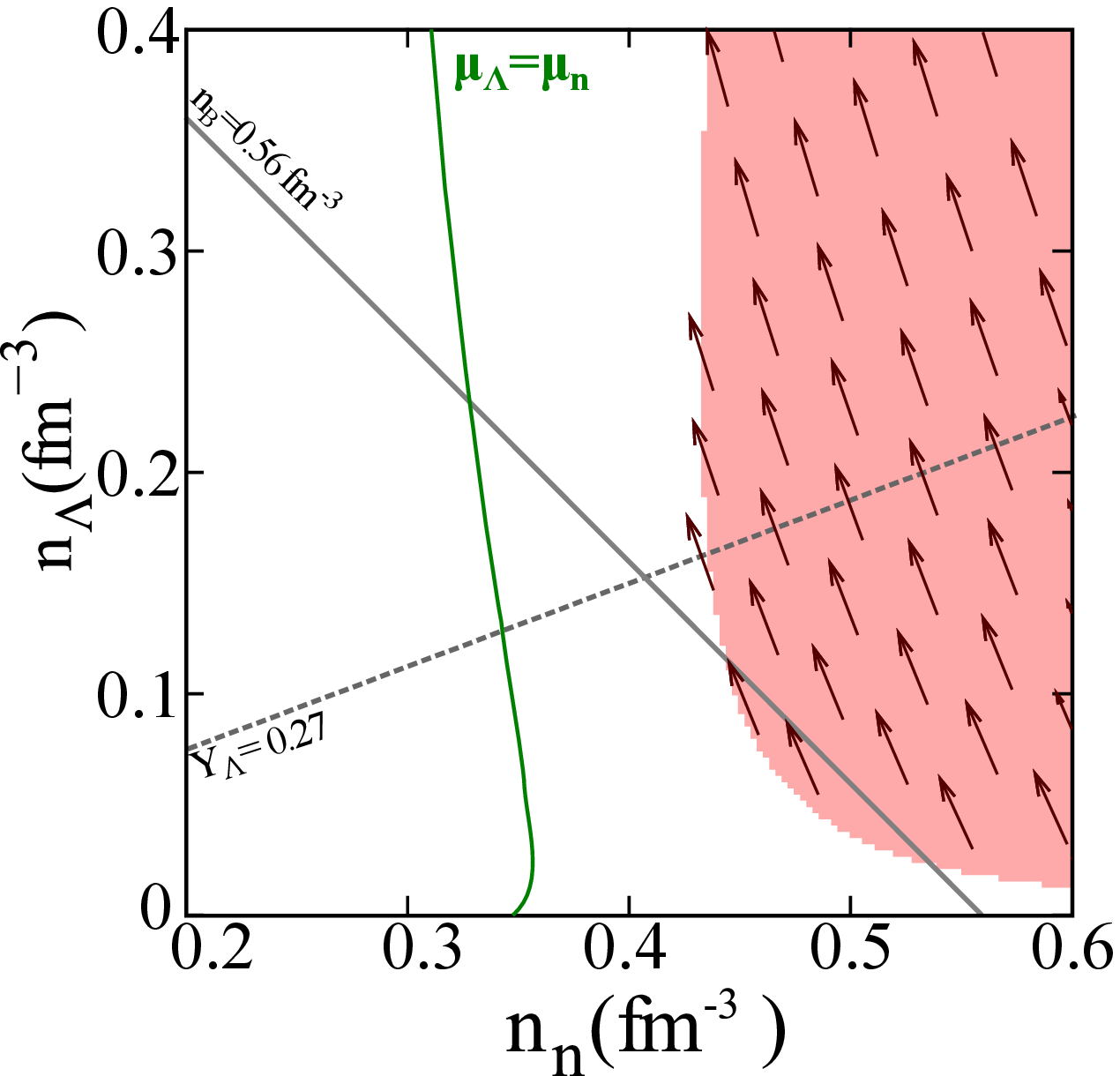}}
\subfloat[]
{\includegraphics[width=0.33\textwidth]{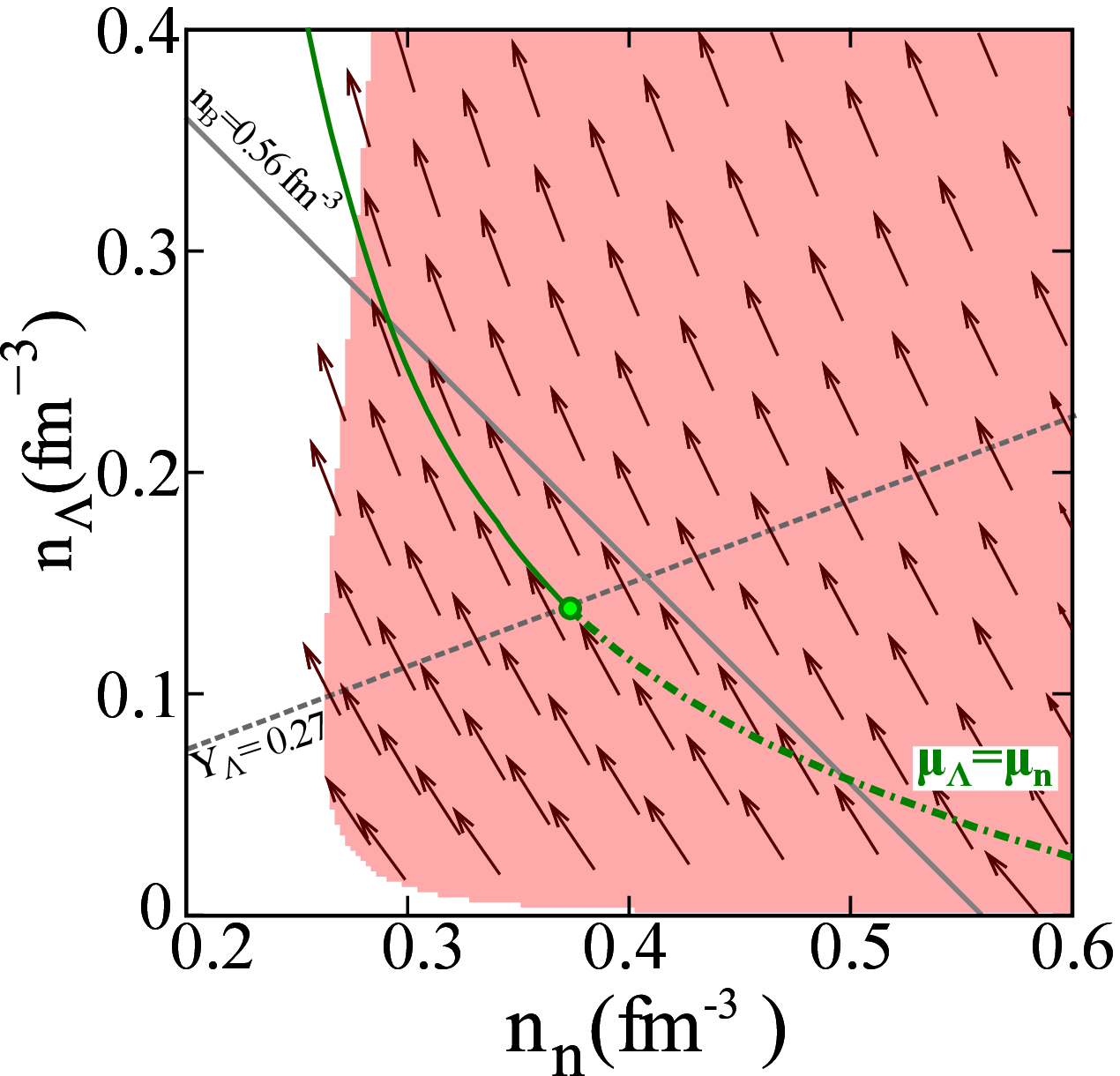}}

\caption{(Color online) Phase diagram of the AFDMC functional with two-body interactions only (left panel), and with two different models I (central panel) and II (right panel)
for the three-body force. The coloured area corresponds to the spinodal region. The arrows indicate the direction of the eigenvectors. The strangeness equilibrium condition $\mu_n=\mu_\Lambda$ is also given, as well as the maximum densities explored in the Monte-Carlo calculation.}
\label{Spinodal ab inito}
\end{figure*}

\paragraph*{}
In  Ref.\cite{abinitio} three different sets of parameters are given, corresponding to a calculation with only two-body forces considered
(hereby called $\Lambda N$), and two extra sets with the addition of  three body forces
that yield two different parameterisations, noted as $\Lambda NN$ (I) and $\Lambda NN$ (II). 
Only the choice $\Lambda NN$ (II) leads to neutron star masses fulfiling the $2 M_\odot$ constraint.
The stability study of the functional Eq.(\ref{pederiva}) is presented in Fig.\ref{Spinodal ab inito}.
It is important to observe that Eq.(\ref{pederiva}) can be plotted for
any arbitrary value of $n_n$ and $n_\Lambda$, but it has been adjusted
to actual AFDMC calculations only in the domain $Y_\Lambda\leq 0.27$
and $n_B\leq 0.56$ fm$^{-3}$. Outside this domain, the functional has
to be considered as an extrapolation.
We can see that a region of instability is always present in some
region of the density space. This is at variance with the RMF results
just presented and qualitatively similar to the case of the
non-relativistic phenomenological BG model 
\cite{Balberg97} studied in \cite{Fran02,Fran04,Fran05,Oertel}. 

\begin{figure}[h]
\begin{tabular}{c}
\subfloat[]
{\includegraphics[width=0.45\textwidth]{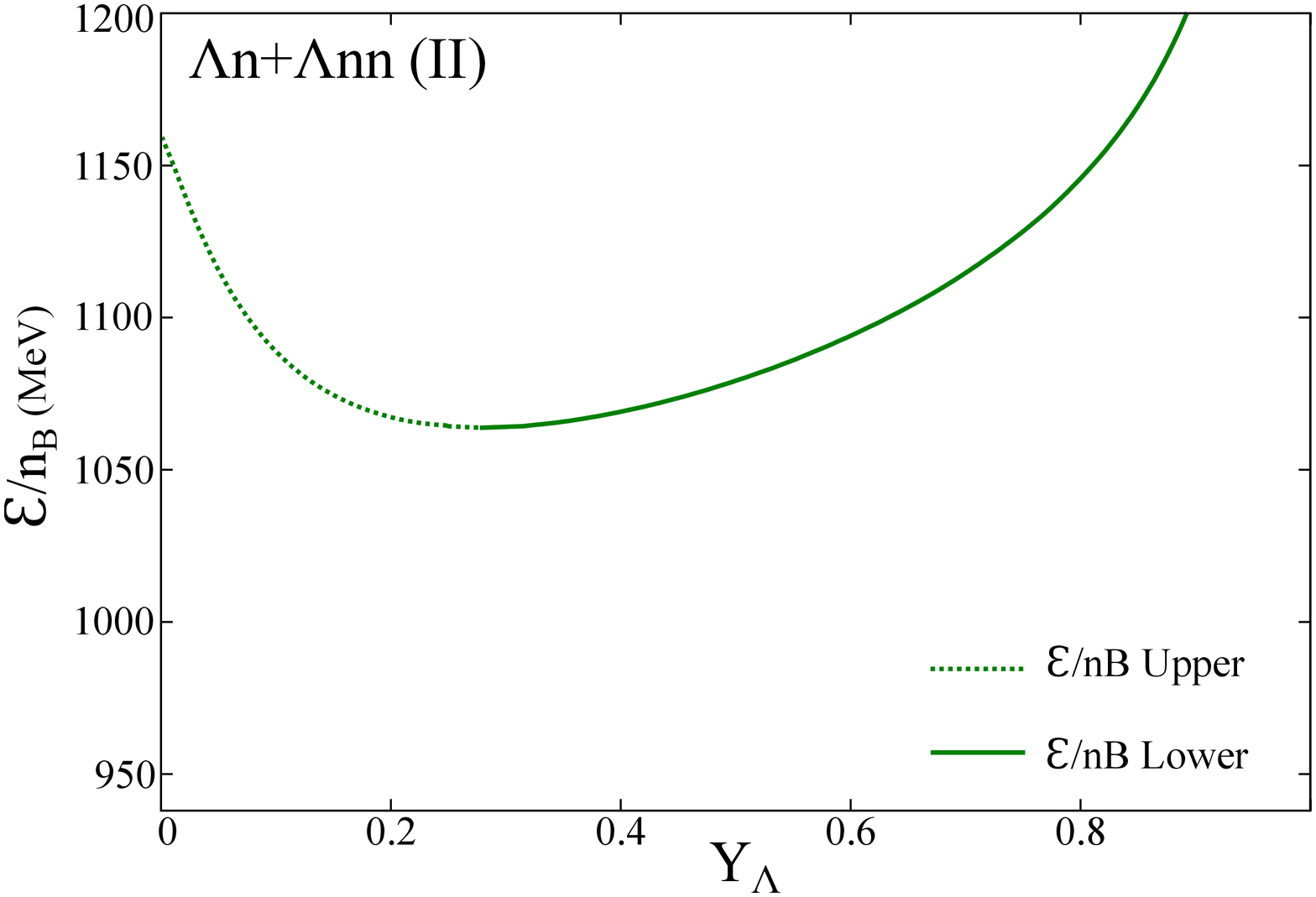}} \\
\subfloat[]
{\includegraphics[width=0.45\textwidth]{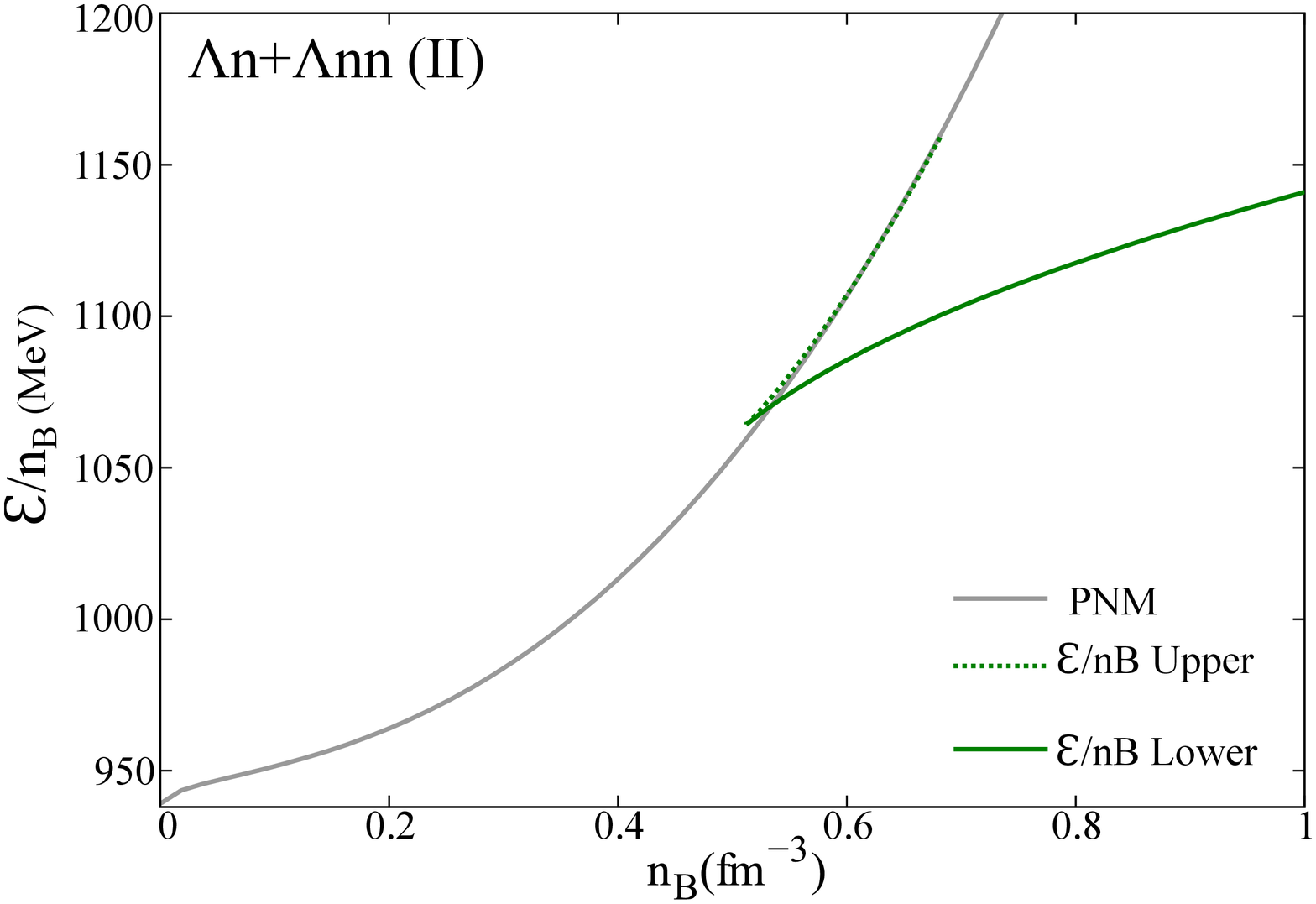}}
\end{tabular}
\caption{(Color online)  Binding energy per baryon as a function of
  the (a) $\Lambda$ fraction and (b) of the baryonic density  within
  the AFDMC functional and model II for the three body force.  
The dotted green lines indicate the unstable branch. The gray line on
the bottom panel corresponds to a calculation of pure neutron matter.}
\label{Bindenergytwobranch}
\end{figure}

\paragraph*{}
In the absence of three-body forces (left part) the instability concerns the whole low neutron density region $0<n_n<0.3$ fm$^-3$. As we can see from the direction of the unstable eigenvectors also given in the figure, this instability  favours the emergence of a dense hyperon-rich phase. This instability zone is also present in the region covered by the microscopic calculations. This means that it cannot be attributed to the extrapolation given by the functional dependence assumed for the energy density, 
and could in principle correspond to a physical phase transition. Still, this effect disappears when including three-body forces. We can therefore conclude that it is most probably an unphysical instability which underlines the importance of three-body forces  (and/or hyperon-hyperon interactions) when studying dense matter. It is important to remark that the strangeness-equilibrium trajectory defined by the chemical potential equality $\mu_n=\mu_\Lambda$ is not affected by this instability. This means that, even if we were facing a spurious instability,  no effect is to be expected in predictions for neutron star matter and the functional can be safely used for neutron star applications. 
With the inclusion of three-body forces, the instability region is pushed towards very high densities.
Within model I (central part of Fig.\ref{Spinodal ab inito}), these densities lie out of the domain covered by the microscopic calculations. This means that the instability might be linked to the extrapolation of Eq.(\ref{pederiva}) out of the domain where it was established.
Again, the instability domain is far from the strangeness equilibrium condition, implying that no consequence is to be expected for applications in neutron star matter.
A more surprising result is obtained within model II (right part of
Fig.\ref{Spinodal ab inito}), which corresponds to the stiffest
equation of state for neutron star matter \cite{abinitio}. For this
model, the instability region extends to a relatively low density
region where in principle the model should be safely applicable.  
Interesting enough, the instability is met at density $n_B\ge 0.51$
fm$^{-3}$ by matter fulfilling the strangeness equilibrium condition
$\mu_n=\mu_\Lambda$, condition which is certainly realized in the
catalyzed matter forming a neutron star.
The peculiar decreasing behaviour of the $\mu_n=\mu_\Lambda$ equilibrium line can be better understood by inspection of Fig.\ref{Bindenergytwobranch},which displays the binding energy behaviour as a function of the $\Lambda$ fraction (left part) and the baryonic density (right part).
We can see that at densities above $n_B= 0.51$ fm$^{-3}$ two different solutions  fulfil
the strangeness equilibrium requirement, and the lower energy solution corresponds to the higher strangeness fraction. This is at the origin of the discontinuous behaviour shown in Fig. \ref{Spinodal ab inito}.

\paragraph*{}
To conclude, the convexity study of the functional  Eq.(\ref{pederiva}) shows that a large region of strangeness driven instability exists in the n-$\Lambda$ system.
We cannot a-priori exclude that it might correspond to a physical phase transition, which could not be seen in the RMF model because of the limitations of the phenomenological approach. However a word of caution is in order. First, no $\Lambda$-$\Lambda$ interaction is considered in \cite{abinitio}, which should probably play an important role at these high densities. Second, the three-body force is purely phenomenological and was adjusted only on hyper-nuclear data, meaning that it might be unrealistic at high density. 
The fact that the instable zone is not upper bound in density indeed signals 
some pathology in the high density extrapolation.
 Finally, the functional Eq.(\ref{pederiva}) was adjusted
only to a limited number of microscopic calculations. It would be
important to know if the Monte-Carlo calculations could be equally
well reproduced by a functional form that does not present any 
convexity anomaly.   It would also be very interesting to see if this behavior is present also in other ab-initio approaches. A complete study of the phase diagram for the BHF functional of ref.\cite{Schulze2016} is in progress.

\subsubsection{Results for np$\Lambda$ stellar matter}

\begin{figure}[h!]
\centering
{\includegraphics[width=0.5\textwidth]{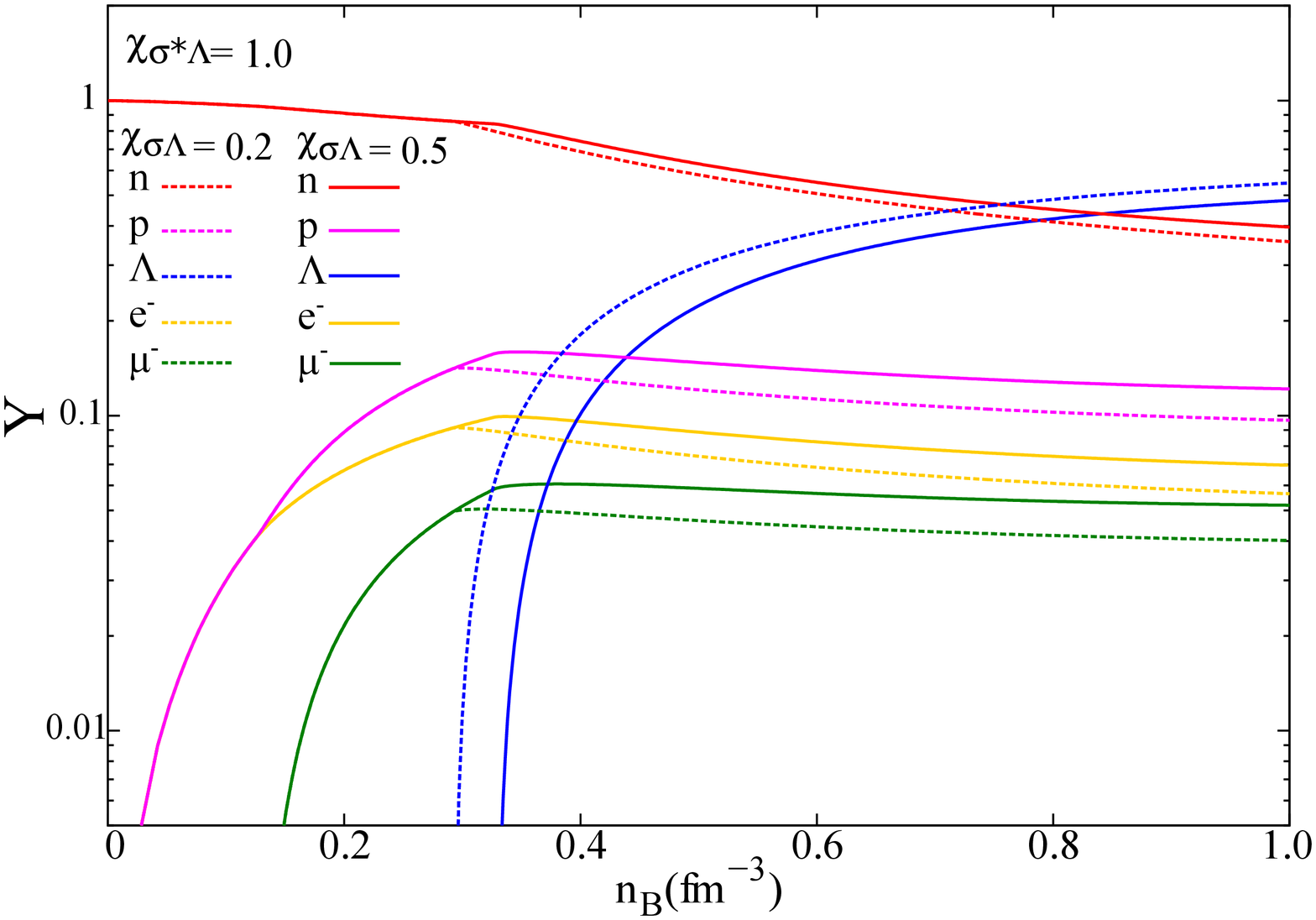}}
\caption{(Color online) Particle relative population in $\left(n~,p~,\Lambda~,e^{-}~,\mu^{-}~\right)$ matter
obtained with $\chi_{\sigma^{\ast} \Lambda}=1.0$ and two values of $\chi_{\sigma \Lambda}$.}
\label{Figure3}
\end{figure}

\begin{figure}[h!]
\centering
{\includegraphics[width=0.5\textwidth]{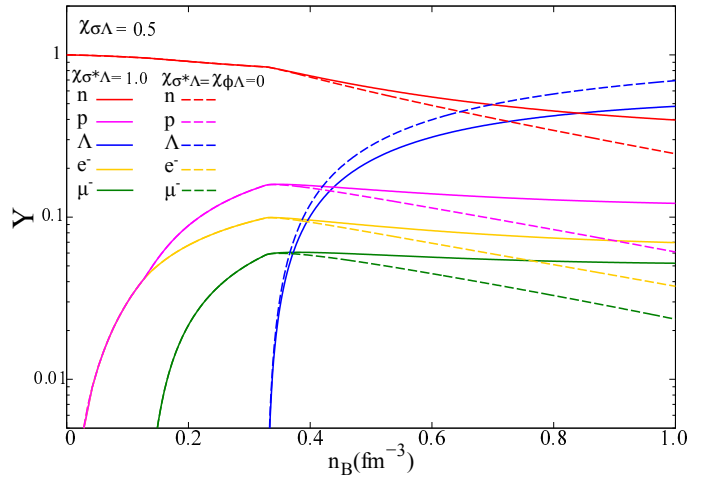}}
\caption{(Color online) Particle relative population in $\left(n~,p~,\Lambda~,e^{-}~,\mu^{-}~\right)$ matter
obtained with $\chi_{\sigma \Lambda}=0.5$ and two values of $\chi_{\sigma^{\ast} \Lambda}$.}
\label{Figure4}
\end{figure}

\paragraph*{}
We are in a position to include protons and leptons and analyse neutron star
matter (charge neutral np$\Lambda$ in chemical equilibrium). Once
more, after the
EOS is obtained, it is used as input to the Tolman-Oppenheimer-Volkoff
(TOV) \cite{tov} equations so that the stellar macroscopic properties
are computed. A reflex
of the previous discussion is the particle population, which we
display in Figures \ref{Figure3} and \ref{Figure4}. While the onset of 
$\Lambda$ is determined by the $\chi_{\sigma \Lambda}$, the amount of 
$\Lambda$ hyperons varies according to the strength the
$\chi_{\sigma^{\ast}  \Lambda}$. In Figure \ref{Figure4} the strange mesons
are turned off so that their role can be better seen.
As already discussed, strangeness can exist in hadronic matter only
if the value of $\chi_{\sigma \Lambda}$ is smaller than 1 (see Table \ref{Table Mmax/M})
because for larger values, either the effective mass of the $\Lambda$s becomes
negative or they simply do not appear due to a very repulsive
potential.

\paragraph*{}
In Table \ref{Table1} we fix two values of the $\chi_{\sigma
  \Lambda}$, both smaller than 1 for the reasons discussed above and
vary  $\chi_{\sigma^{\ast} \Lambda}$. We then calculate the
corresponding maximum mass, radius, central energy density, central
baryonic density, central fraction of $\Lambda$ particles and the
density related to the onset of $\Lambda$s.   As already observed from Fig.\ref{Grid} and from the results obtained in \cite{low-density}, 
because of the positive correlation between the scalar and the vector couplings 
imposed by the constraints,  the increase of the $\chi_{\sigma \Lambda}$ ($\chi_{\sigma^{\ast}    \Lambda}$) makes the potential $U^N_\Lambda$ ($U^\Lambda_\Lambda$)  deeper at low densities (due to the scalar meson attraction)  but also steeper at high densities (due to the correlated vector meson repulsion). For this reason, higher values of the scalar couplings lead to stiffer EOS,
  resulting in more massive stars, with a smaller content of $\Lambda$
  hyperons.  
In all cases seen in Table \ref{Table1}, the central density is larger than the
  density where $\Lambda$ particles appear and, even with the presence
  of these particles, very massive stars can be attained. Another
  point worth mentioning is that the smaller the value of the
  $\chi_{\sigma \Lambda}$, the larger the variation of the maximum
  stellar mass due to different values of $\chi_{\sigma^{\ast}
    \Lambda}$.

\begin{table}
\begin{tabular}{ccccccc}
\hline
\multicolumn{7}{c}{ $\chi_{\sigma \Lambda}=0.2$} \\
\hline
 $\chi_{\sigma^{\ast} \Lambda}$ &$M_{\text{max}}$ (M$_{\odot}$) & R (km) &  $\epsilon_c$ (fm$^{-4}$) & $n_c$(fm$^{-3}$)  & $Y_{\Lambda}^{c}$    & $n^{\Lambda}_{\text{onset}}$(fm$^{-3}$)  \\
$0.5$                           & $-$                    & $-$     & $-$                       & $-$               &$-$            & $-$	\\
$1.0$                           & $1.44$                 & $9.50$  & $11.02$                   & $1.73$            &$0.63$         & $0.29$	\\
$3.5$                           & $2.13$                 & $10.97$ & $6.96$                    & $1.04$            &$0.16$         & $0.29$	\\
$4.0$                           & $2.17$                 & $11.14$ & $6.71$                    & $1.01$            &$0.13$         & $0.29$	\\
\hline
\multicolumn{7}{c}{  $\chi_{\sigma \Lambda}=0.8$} \\
\hline
 $\chi_{\sigma^{\ast} \Lambda}$ &$M_{\text{max}}$ (M$_{\odot}$) & R (km) &  $\epsilon_c$ (fm$^{-4}$) & $n_c$(fm$^{-3}$)  & $Y_{\Lambda}^{c}$    & $n^{\Lambda}_{\text{onset}}$(fm$^{-3}$)  \\ 
$0.5$                           & $2.28$                 & $11.91$  & $5.72$                   & $0.88$            &$0.25$         & $0.39$	\\
$1.0$                           & $2.29$                 & $11.87$  & $5.76$                   & $0.88$            &$0.21$         & $0.39$	\\
$3.5$                           & $2.33$                 & $11.85$  & $5.77$                   & $0.88$            &$0.06$         & $0.39$	\\
$4.0$                           & $2.34$                 & $11.87$  & $5.75$                   & $0.87$            &$0.05$         & $0.39$	\\
\hline
\end{tabular}
\caption{Stellar properties obtained from
  $\left(n~,p~,\Lambda~,e^{-}~,\mu^{-}~\right)$ matter with different
  values of the coupling constants. The symbol $-$ means that the
  maximum stellar mass is smaller than 1.44 $M_\odot$ and the results
  are disregarded.}
\label{Table1}
\end{table}

\subsection{Eight lightest baryons included} 

\paragraph*{}
To start our considerations, we go back to the literature and choose
different possible ways to fix the meson-hyperon coupling constants when
the 8 lightest baryons are taken into account. 
The simplest possibility \cite{GlennBook} is to employ for all hyperons the same couplings established for the $\Lambda$, $\chi_{\sigma  H}=\chi_{\sigma  \Lambda}$, $\chi_{\sigma^{\ast} H}=\chi_{\sigma^{\ast} \Lambda}$, $\chi_{\omega  H}=\chi_{\omega  \Lambda}$, $\chi_{\phi  H}=\chi_{\phi  \Lambda}$.
 This hypothesis satisfies  the inequalities
$g_{\sigma N} \le \frac{1}{2} (3 g_{\sigma \Lambda} + g_{\sigma
  \Sigma}) \le 2 g_{\sigma N} $ imposed by the SU(3)  symmetry
\cite{Colucci2013}, and has the advantage of avoiding the introduction
of other unconstrained coupling constants.
Much stronger hypothesis, where  $\chi_{\sigma  H}=\chi_{\omega H}$,
also without formal justifications, can be found in the literature  
\cite{boguta_1981,rufa_1990}, with different purposes. Nevertheless,
other more realistic possibilities are presented in Table
\ref{Table2}:
in set A, the vector meson coupling constants are fixed according to
SU(6) symmetry, but no strange mesons are included; in set B, SU(6) is
used also for the strange mesons, i.e.,
$2 \chi_{\phi \Lambda}= 2\chi_{\phi  \Sigma}=\chi_{\phi \Xi} = 2 \sqrt 2/3$ and in set C, we have increased the values
of the $\chi_{\sigma^{\ast} H}$ as 
$2 \chi_{\phi \Lambda}= 2\chi_{\phi  \Sigma}=\chi_{\phi \Xi} = 4 \sqrt
2/3$ because we are aware (from the analyses done in
\cite{low-density} ) that the increase of this coupling constant makes
the potential more repulsive, resulting in a stiffer EOS. 
To obtain the other hyperon potentials from the
$U_{\Lambda}^{\Lambda}$, we have assumed that 
$U_{\Sigma}^{\Sigma}(n_0)= U_{\Lambda}^{\Lambda}(n_0/2)$ and
$U_{\Xi}^{\Xi}(n_0)= U_{\Lambda}^{\Lambda}(n_0/2)$, as proposed a long
time ago based on simple strange quark counting arguments 
\cite{schaffner94} and still used nowadays \cite{Oertel}. 
It is important to observe that these parameter sets respect the
constraints discussed in 
Section \ref{Formalism} in the $\Lambda$ sector, see Fig.\ref{Grid} above.
Using these 3 sets to compute the EOS, the associated stellar
properties are displayed in Table \ref{Table3}, where $Y_H$ stands for
the fraction of all hyperons. Only set C results in a star with mass
close to 2 $M_\odot$ and the radii are always a bit large, if we assume that
the recent analysis that predict radii in the range $10.1-11.1$ Km
\cite{ozel} and $12.1 \pm 1.1$ Km \cite{haensel} for 1.4 $M_\odot$
stars  are to be taken as reliable. Notice that 1.4 $M_\odot$  stars
always have radii larger than the $M_{max}$ stars.

\begin{table}[h!]
\begin{center}
\begin{tabular}{c|cccccc}
Set&\multicolumn{5}{c}{Parameters} \\
\hline
\hline
-&$\chi_{\sigma \Lambda}$       & $\chi_{\omega \Lambda}$&$\chi_{\sigma^{\ast} \Lambda}$&$\chi_{\phi \Lambda}$     &$\chi_{\rho \Lambda}$    \\ 
A&$0.61$                        & $2/3$                 & $-$                          & $-$                       & $0.0$ \\
B&$0.61$                        & $2/3$                 & $0.06$                       & $1\times\sqrt{2}/3$       & $0.0$       \\
C&$0.61$                        & $2/3$                 & $0.94$                       & $2\times \sqrt{2}/3$      & $0.0$            \\  
\hline
\hline
-&$\chi_{\sigma \Sigma}$        & $\chi_{\omega \Sigma}$ &$\chi_{\sigma^{\ast} \Sigma}$ &$\chi_{\phi \Sigma}$     &$\chi_{\rho \Sigma}$    \\ 
A&$0.40$                        & $2/3$                 & $-$                          & $-$                      & $1.0$              \\
B&$0.40$                        & $2/3$                 & $0.860$                      & $1\times\sqrt{2}/3$      & $1.0$             \\
C&$0.40$                        & $2/3$                 & $1.280$                      & $2\times\sqrt{2}/3$      & $1.0$      \\
\hline
\hline
-&$\chi_{\sigma \Xi}$           & $\chi_{\omega \Xi}$    &$\chi_{\sigma^{\ast} \Xi}$   &$\chi_{\phi \Xi}$         &$\chi_{\rho \Xi}$  \\ 
A&$0.32$                        & $1/3$                & $-$                           & $-$                      & $2.0$              \\
B&$0.32$                        & $1/3$                & $1.050$                       & $1\times2\sqrt{2}/3$     & $2.0$     \\
C&$0.32$                        & $1/3$                & $2.150$                       & $2\times2\sqrt{2}/3$     & $2.0$     \\ 
\hline 
\end{tabular}
\caption{Coupling constants}
\label{Table2}
\end{center}
\end{table}

\begin{table}[h!]
\begin{center}
\begin{tabular}{cccccccc}
\hline 
Set&$M_{\text{max}}$ (M$_{\odot}$) & R (km)    &  $\epsilon_c$ (fm$^{-4}$) & $n_c$(fm$^{-3}$)  & $Y_\Lambda$  & $Y_H$   & $n^{\Lambda}_{\text{onset}}$(fm$^{-3}$)\\
A  &$1.89$                         & $12.78$   & $4.58$                    & $0.78$            &$0.33$        &$0.57$   & $0.34$                               \\
B  &$1.91$                         & $12.58$   & $4.87$                    & $0.82$            &$0.04$        &$0.37$   & $0.34$                               \\
C  &$1.98$                         & $11.32$   & $6.67$                    & $1.04$            &$0.11$        &$0.39$   & $0.34$                               \\
\hline 
\end{tabular}
\caption{Stellar properties with 8 baryons and zero temperature}
\label{Table3}
\end{center}
\end{table}

\begin{figure}[h]
\begin{tabular}{c}
\subfloat[]
{\includegraphics[width=0.5\textwidth]{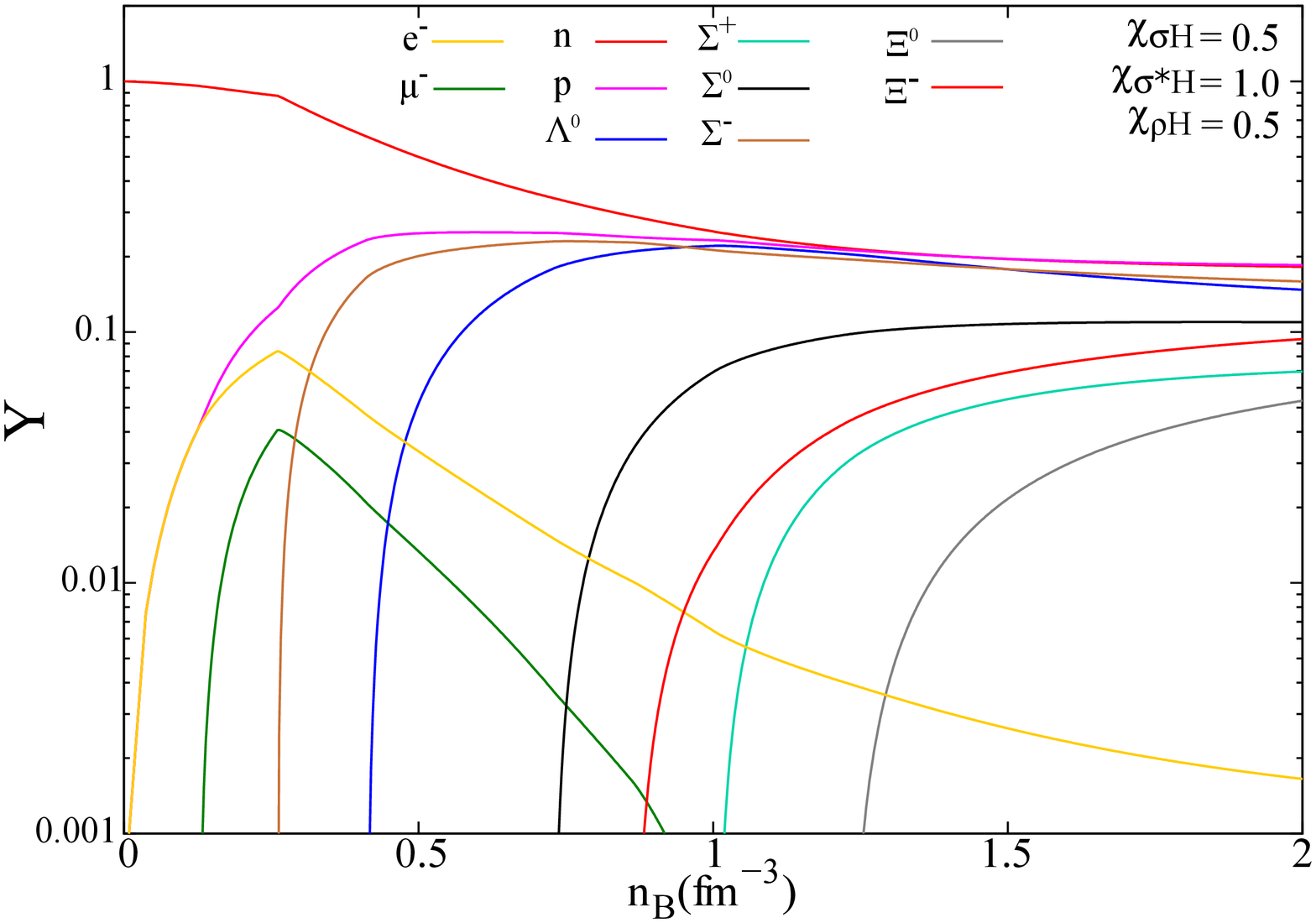}}\\
\subfloat[]
{\includegraphics[width=0.5\textwidth]{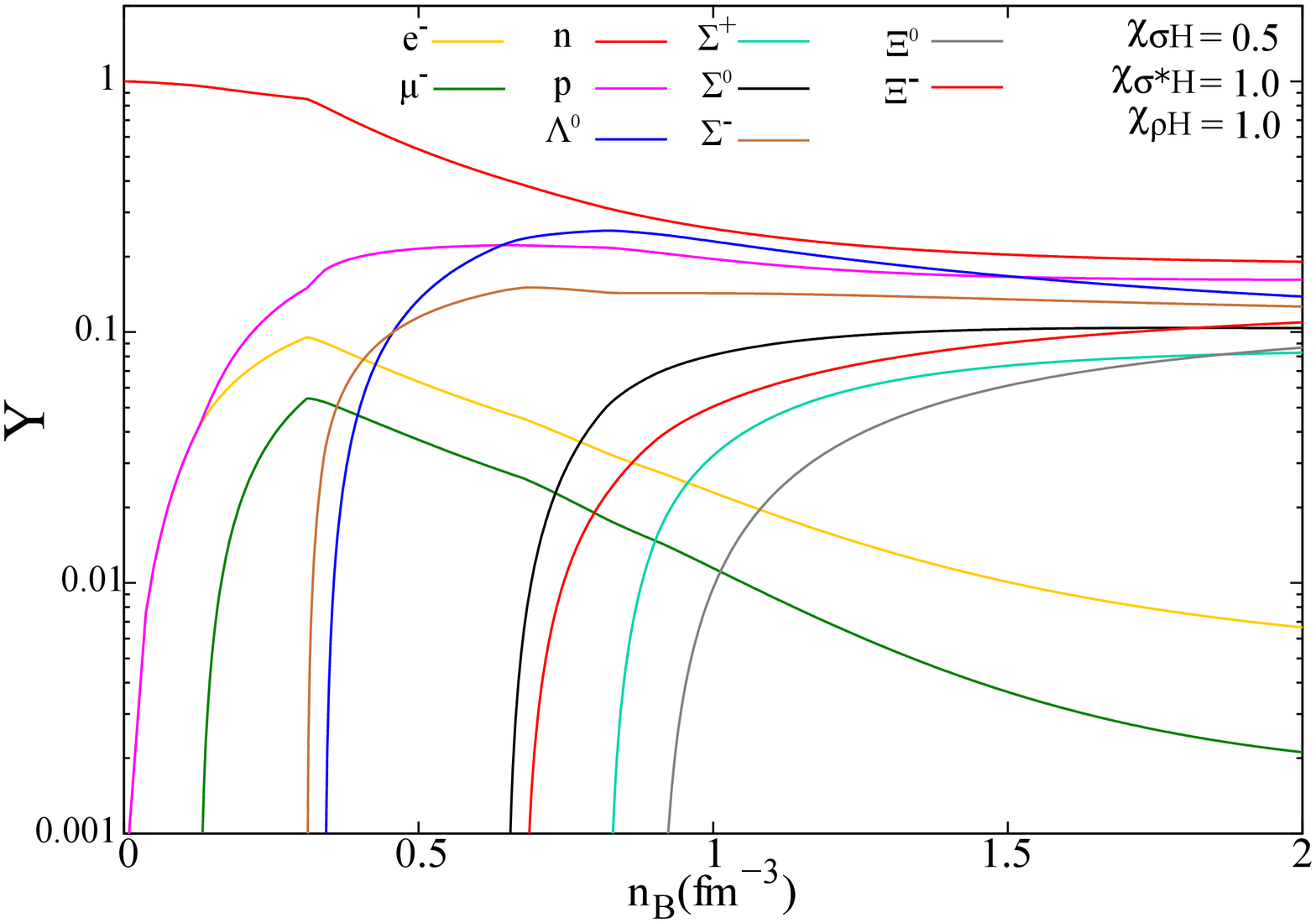}}
\end{tabular}
\caption{(Color online) Particle relative population in
$\left(\text{octet},~e^{-},~\mu^{-}\right)$ with $\chi_{\sigma H}=0.5$, $\chi_{\sigma^{\ast} H}=1.0$ and 
(a)  $\chi_{\rho H}=0.5$, (b), $\chi_{\rho H}=1.0$.}
\label{Figure5}
\end{figure}

\paragraph*{}
We now return to our choice of coupling constants and 
try to identify the individual roles played by the $\rho$ and the strange
  mesons. The effect of the $\chi_{\rho H}$ parameter in the particle population 
can only be singled out within the simplifying hypothesis of equal  
  couplings of the different hyperons with the mesons, i.e, 
$\chi_{\sigma  \Lambda} = \chi_{\sigma \Sigma}=\chi_{\sigma \Xi} = \chi_{\sigma H}$. 
In Figure \ref{Figure5}  $\chi_{\sigma H}$ and $\chi_{\sigma^{\ast} H}$ are
fixed and   $\chi_{\rho H}$ is fixed as 0.5 and 1.0, for $H=\Sigma,\Xi$. It is visually
clear that the constituents can change quite drastically. In this
case, the onset of
$\Lambda$s is at a lower density than the onset of $\Sigma^-$ only for
$\chi_{\rho H}$ around 1.5 or larger (within the GM1 parameterisation,
used in the present work), as already discussed. 

\paragraph*{}
Having in mind that a larger $\chi_{\rho H}$ makes the EOS
stiffer, we fix it at 1.0 and perform the same analyses done for np$\Lambda$
matter, i.e., we choose two values for $\chi_{\sigma H}$ and, for each
of them, vary $\chi_{\sigma^{\ast} H}$. The results are shown in Table
\ref{Table4}. One can see that the same pattern found before with
$\Lambda$s only repeats itself when the whole octet is allowed to appear:
the onset of hyperons is determined by the $\chi_{\sigma H}$ because the
larger its value, the larger the repulsion felt by them and 
consequently, the latter in density is their onset, resulting in stiffer EOS.
The $\chi_{\sigma^{\ast}  H}$ determines the amount of hyperons in the
system. Massive stars can be obtained with a fair amount of
strangeness and radii in the interval foreseen in \cite{haensel} can
also be found. 
This qualitative result is in agreement with the findings of 
\cite{Oertel}, though the hypernuclear constraint $U_{\Lambda}^{\Lambda}\approx -0.67$ MeV was not considered in that paper,
considerably increasing the dimension of the parameter space in that work.

\begin{table}[h!]
\begin{center}
\begin{tabular}{cccccccc}
\hline
\multicolumn{8}{c}{$\chi_{\sigma H}=0.2$  }\\
\hline
 $\chi_{\sigma^{\ast} H}$ &$M_{\text{max}}$ (M$_{\odot}$) & R (km) &  $\epsilon_c$ (fm$^{-4}$) & $n_c$(fm$^{-3}$)  & $Y_{\Lambda}^{c}$ & $Y_{H}^{c}$    & $n^{\Lambda}_{\text{onset}}$(fm$^{-3}$)  \\
$0.5$                           & $-$                    & $-$      & $-$                      & $-$               &$-$                &$-$             & $-$	\\
$1.0$                           & $-$                    & $-$      & $-$                      & $-$               &$-$                &$-$             & $-$	\\
$3.5$                           & $2.09$                 & $10.73$  & $7.35$                   & $1.09$            &$0.02$             &$0.18$          & $0.30$	\\
$4.0$                           & $2.14$                 & $10.93$  & $7.00$                   & $1.05$            &$0.01$             &$0.15$          & $0.30$	\\
\hline
\multicolumn{8}{c}{NLWM - $\chi_{\sigma H}=0.8$ } \\
\hline
 $\chi_{\sigma^{\ast} H}$ &$M_{\text{max}}$ (M$_{\odot}$) & R (km) &  $\epsilon_c$ (fm$^{-4}$) & $n_c$(fm$^{-3}$)  & $Y_{\Lambda}^{c}$   & $Y_{H}^{c}$   & $n^{\Lambda}_{\text{onset}}$(fm$^{-3}$)  \\
$0.5$                           & $2.24$                 & $11.83$  & $5.84$                   & $0.90$            &$0.14$         &$0.33$         & $0.42$	\\
$1.0$                           & $2.25$                 & $11.78$  & $5.90$                   & $0.90$            &$0.11$         &$0.26$         & $0.42$	\\
$3.5$                           & $2.32$                 & $11.78$  & $5.86$                   & $0.89$            &$0.00$         &$0.08$         & $0.42$	\\
$4.0$                           & $2.32$                 & $11.80$  & $5.84$                   & $0.88$            &$0.00$         &$0.06$         & $0.42$	\\
\hline
\end{tabular}
\caption{Stellar properties with 8 baryons and zero temperature for
  $\chi_{\rho H}=1.0$  for $H=\Sigma,\Xi$ and  different values of the other coupling constants. The symbol $-$ means that the
  maximum stellar mass is smaller than 1.44 $M_\odot$ and the results
  are disregarded.}
\label{Table4}
 \end{center}
\end{table}

\paragraph*{}
In the stellar evolution, three snapshots  of the time evolution of a star in
its first minutes of life are usually considered and they are
simulated though different entropies per particle:
\begin{itemize}
\item $s/n_B=1$, $Y_l=0.4$,
\item $s/n_B=2$, $\mu_{\nu_l}=0$,
\item $s/n_B=0$, $\mu_{\nu_l}=0$,
\end{itemize}
where the entropy per particle (baryon) can be calculated through the thermodynamical expression
\begin{eqnarray}
\frac{S}{A}= \frac{s}{n_B}=\frac{\varepsilon+  p -\sum_j \mu_j n_j -\sum_l \mu_l n_l}{T n_B}.
\end{eqnarray}
At first, the star is relatively
warm (represented by fixed entropy per particle) and has a large number of
trapped neutrinos (represented by fixed lepton fraction). As the trapped 
neutrinos diffuse, they heat up the star
\cite{Prakash:1996xs}. Finally, the star is completely deleptonised and
can be considered cold. In the present analysis, we disregard the
neutrinos because, although they are very important in studies
involving transport equations and star cooling processes, they play a minor role in the
determination of the maximum stellar mass, its radius and other
stellar properties, our main concern here. The results obtained for a
fixed value of $\chi_{\sigma H}$ and different choices of
$\chi_{\sigma^{\ast} H}$ are displayed in Table \ref{Table5}, where we
also show the value of the star central temperature ($T_c$). Again
$\chi_{\rho H} = 1$ for $H=\Sigma,\Xi$. The onset of $\Lambda$s take place at very low
densities, although with very small fractions at first. This can be easily
understood because temperature helps the appearance of heavier
particles. As an obvious result, we can see that the star central density increases
with the entropy, but the values are not too high.
Notice that the inclusion of the strange mesons makes the EOS stiffer,
as already pointed out in many other works both at zero \cite{Weissenborn11b,
Weissenborn11c, Luiz} and finite temperature or with fixed entropies \cite{Cavagnoli2} and this fact
does not depend on the value of the coupling constants, which only
control how stiffer the EOS becomes. As far as the radii are
concerned, we can see that they decrease with the inclusion of the
lightest 8 baryons in comparison with matter containing $\Lambda$s
only (compare Tables \ref{Table1} and \ref{Table4}). Massive stars can
be obtained with a reasonable amount of hyperons (up to $Y_H=0.33$).

\begin{table*}[t]
\begin{center}
\begin{tabular}{cccccccc}
\hline 
\multicolumn{8}{c}{$\chi_{\sigma H}=0.7$ e $\chi_{\sigma^{\ast} H}=\chi_{\phi H}=0.0$ } \\
\hline
$s/n_{B}$ &$M_{\text{max}}$ (M$_{\odot}$) & R (km)    &  $\epsilon_c$ (fm$^{-4}$) & $n_c$(fm$^{-3}$)  & $Y_\Lambda$   & $n^{\Lambda}_{\text{onset}}$(fm$^{-3}$) &   $T_{c}$ (MeV)\\ 
$0$  &$2.01$                         & $12.10$   & $5.55$                    & $0.90$            &$0.31$         & $0.38$	                              &    $0.00$\\
$1$  &$2.01$                         & $12.14$   & $5.55$                    & $0.89$            &$0.29$         & $-$	                                      &    $18.24$\\
$2$  &$2.00$                         & $12.19$   & $5.55$                    & $0.88$            &$0.25$         & $-$	                                      &    $40.52$\\
\hline 
\multicolumn{8}{c}{$\chi_{\sigma H}=0.7$ e $\chi_{\sigma^{\ast} H}=1.0$ } \\
\hline
$s/n_{B}$ &$M_{\text{max}}$ (M$_{\odot}$) & R (km)    &  $\epsilon_c$ (fm$^{-4}$) & $n_c$(fm$^{-3}$)  & $Y_\Lambda$   & $n^{\Lambda}_{\text{Lim}}$(fm$^{-3}$) &   $T_{c}$ (MeV)\\ 
$0$  &$2.15$                         & $11.66$   & $6.10$                    & $0.94$            &$0.16$         & $0.38$	                              &    $0.00$\\
$1$  &$2.15$                         & $11.66$   & $5.55$                    & $0.94$            &$0.17$         & $-$	                                      &    $24.13$\\
$2$  &$2.16$                         & $11.78$   & $5.55$                    & $0.91$            &$0.15$         & $-$	                                      &    $50.65$\\
\hline 
\multicolumn{8}{c}{$\chi_{\sigma H}=0.7$ e $\chi_{\sigma^{\ast} H}=1.5$ } \\
\hline
$s/n_{B}$ & $M_{\text{max}}$ (M$_{\odot}$) & R (km)    &  $\epsilon_c$ (fm$^{-4}$) & $n_c$(fm$^{-3}$)  & $Y_\Lambda$   & $n^{\Lambda}_{\text{Lim}}$(fm$^{-3}$) & $T_{c}$ (MeV)\\ 
$0$  & $2.18$                         & $11.55$   & $6.21$                    & $0.95$            &$0.09$         & $0.38$	                           &    $0.00$\\
$1$  &$2.20$                          & $11.58$   & $6.22$                    & $0.94$            &$0.11$         & $-$	                                   &    $26.15$\\
$2$  &$2.20$                          & $11.75$   & $6.08$                    & $0.91$            &$0.10$         & $-$	                                   &    $53.67$\\
\hline 
\end{tabular}
\caption{Stellar properties for three snapshots of the star evolution
  with $\chi_{\rho H}=1.0$ ( for $H=\Sigma,\Xi$) and different values for $\chi_{\sigma
    H}=0.7$ e $\chi_{\sigma^{\ast} H}=1.0$ .
 The symbol $-$ means that the density is very low.}
\label{Table5}
\end{center}
\end{table*}

\section{Final Remarks}

\paragraph*{}
We have shown that after a large region of possible coupling constants was spanned, no
instabilities were found, indicating that no strange phase transition 
is expected to appear inside neutron stars at least in the framework
of the RMF model. 
This strong conclusion could be found thanks to the restriction 
of the parameter space due to the constraints extracted from hypernuclear data.
This is in contrast with what was found with
non-relativistic models and also in \cite{Oertel}, where a spinodal instability was obtained at the
onset of hyperons, but only as far as very high coupling constants,
not compatible with present constraints from hypernuclear physics,
were used. This fact reinforces the idea that a phase transition at
high density stellar matter would require the presence of
quarks. Hence, hybrid stars containing a quark core should not be discarded.
 
\paragraph*{}
Moreover, we saw that hyperons can coexist with nucleons giving
  rise to massive stars bearing a radii of the order of 10-12 Km,
  compatible with 1.4 $M_\odot$ stars with radii approximately 10\%
  larger, values which agree with a recent calculation \cite{haensel}. 
When the  full octet is allowed to exist, the radius tend to decrease. Massive
  stars can be obtained with a large amount of hyperons as far as
  strange mesons are included in the formalism, result that helps us to
  shed some light to the hyperon puzzle. It is never too much to say
  again that all these results are parameter and model dependent, what
  reinforces the need of further contraints at high densities. 

\paragraph*{}
As a final remark, we claim that it would be interesting to confront the values of the coupling
constants allowed by the analyses of phenomenological potentials  with
the values obtained from the assumption of an SU(3) symmetry done in
\cite{Luiz} when the scalar mesons are included, a work already in progress. 

\begin{acknowledgments}
 This work was partially supported by CAPES/COFECUB project 853/15 and
 CNPq under grant 300602/2009-0.
\end{acknowledgments}

\end{document}